\documentclass[12pt]{article}
\usepackage{graphics}
\usepackage{epsf}
\begin{document}

\newcommand{\be}{\begin{equation}}
\newcommand{\beq}{\begin{equation}}
\newcommand{\eeq}{\end{equation}}
\newcommand{\ee}{\end{equation}}

\newcommand{\beqn}{\begin{eqnarray}}
\newcommand{\eeqn}{\end{eqnarray}}
\newcommand{\bea}{\begin{eqnarray}}
\newcommand{\ena}{\end{eqnarray}}
\newcommand{\ra}{\rightarrow}
\newcommand{\susy}{{{\cal SUSY}$\;$}}
\newcommand{\su}{$ SU(2) \times U(1)\,$}

\newcommand{\gag}{$\gamma \gamma$ }
\newcommand{\gam}{\gamma \gamma }
\def\W{{\mbox{\boldmath $W$}}}
\def\B{{\mbox{\boldmath $B$}}}
\newcommand{\np}{Nucl.\,Phys.\,}
\newcommand{\pl}{Phys.\,Lett.\,}
\newcommand{\pr}{Phys.\,Rev.\,}
\newcommand{\prl}{Phys.\,Rev.\,Lett.\,}
\newcommand{\prep}{Phys.\,Rep.\,}
\newcommand{\zp}{Z.\,Phys.\,}
\newcommand{\sovjnp}{{\em Sov.\ J.\ Nucl.\ Phys.\ }}
\newcommand{\nuclinst}{{\em Nucl.\ Instrum.\ Meth.\ }}
\newcommand{\annp}{{\em Ann.\ Phys.\ }}
\newcommand{\intjmp}{{\em Int.\ J.\ of Mod.\  Phys.\ }}

\newcommand{\eps}{\epsilon}
\newcommand{\mw}{M_{W}}
\newcommand{\mww}{M_{W}^{2}}
\newcommand{\mwmw}{M_{W}^{2}}
\newcommand{\mhmh}{M_{H}^2}
\newcommand{\mz}{M_{Z}}
\newcommand{\mzz}{M_{Z}^{2}}

\newcommand{\lra}{\leftrightarrow}
\newcommand{\tr}{{\rm Tr}}
\def\ls1{{\not l}_1}
\newcommand{\cms}{centre-of-mass\hspace*{.1cm}}

\newcommand{\dkg}{\Delta \kappa_{\gamma}}
\newcommand{\dkz}{\Delta \kappa_{Z}}
\newcommand{\dz}{\delta_{Z}}
\newcommand{\dgz}{\Delta g^{1}_{Z}}
\newcommand{\dgzt}{$\Delta g^{1}_{Z}\;$}
\newcommand{\la}{\lambda}
\newcommand{\lag}{\lambda_{\gamma}}
\newcommand{\lambdae}{\lambda_{e}}
\newcommand{\laz}{\lambda_{Z}}
\newcommand{\lnl}{L_{9L}}
\newcommand{\lnr}{L_{9R}}
\newcommand{\lt}{L_{10}}
\newcommand{\lu}{L_{1}}
\newcommand{\ld}{L_{2}}
\newcommand{\cw}{\cos\theta_W}
\newcommand{\sw}{\sin\theta_W}
\newcommand{\tw}{\tan\theta_W}
\def\cww{\cos^2\theta_W}
\def\sww{\sin^2\theta_W}
\def\tww{\tan^2\theta_W}

\newcommand{\epm}{$e^{+} e^{-}\;$}
\newcommand{\epemt}{$e^{+} e^{-}\;$}
\newcommand{\epem}{e^{+} e^{-}\;}
\newcommand{\ememt}{$e^{-} e^{-}\;$}
\newcommand{\emem}{e^{-} e^{-}\;}
\newcommand{\eeww}{e^{+} e^{-} \ra W^+ W^- \;}
\newcommand{\eewwt}{$e^{+} e^{-} \ra W^+ W^- \;$}
\newcommand{\epemww}{e^{+} e^{-} \ra W^+ W^- }
\newcommand{\epemwwt}{$e^{+} e^{-} \ra W^+ W^- \;$}
\newcommand{\eennhht}{$e^{+} e^{-} \ra \nu_e \bar \nu_e HH\;$}
\newcommand{\eennhh}{e^{+} e^{-} \ra \nu_e \bar \nu_e HH\;}
\newcommand{\ppwg}{p p \ra W \gamma}
\newcommand{\wwhh}{W^+ W^- \ra HH\;}
\newcommand{\wwhht}{$W^+ W^- \ra HH\;$}
\newcommand{\ppwz}{pp \ra W Z}
\newcommand{\ppwgt}{$p p \ra W \gamma \;$}
\newcommand{\ppwzt}{$pp \ra W Z \;$}
\newcommand{\gamgamt}{$\gamma \gamma \;$}
\newcommand{\gamgam}{\gamma \gamma \;}
\newcommand{\egamt}{$e \gamma \;$}
\newcommand{\egam}{e \gamma \;}
\newcommand{\gamgamwwt}{$\gamma \gamma \ra W^+ W^- \;$}
\newcommand{\gamgamwwht}{$\gamma \gamma \ra W^+ W^- H \;$}
\newcommand{\gamgamwwh}{\gamma \gamma \ra W^+ W^- H \;}
\newcommand{\gamgamwwhht}{$\gamma \gamma \ra W^+ W^- H H\;$}
\newcommand{\gamgamwwhh}{\gamma \gamma \ra W^+ W^- H H\;}
\newcommand{\ggww}{\gamma \gamma \ra W^+ W^-}
\newcommand{\ggwwt}{$\gamma \gamma \ra W^+ W^- \;$}
\newcommand{\ggwwht}{$\gamma \gamma \ra W^+ W^- H \;$}
\newcommand{\ggwwh}{\gamma \gamma \ra W^+ W^- H \;}
\newcommand{\ggwwhht}{$\gamma \gamma \ra W^+ W^- H H\;$}
\newcommand{\ggwwhh}{\gamma \gamma \ra W^+ W^- H H\;}
\newcommand{\ggwwz}{\gamma \gamma \ra W^+ W^- Z\;}
\newcommand{\ggwwzt}{$\gamma \gamma \ra W^+ W^- Z\;$}
\def\smx{{\cal{S}}{\cal{M}}}

\newcommand{\ptu}{p_{1\bot}}
\newcommand{\vecptu}{\vec{p}_{1\bot}}
\newcommand{\ptd}{p_{2\bot}}
\newcommand{\vecptd}{\vec{p}_{2\bot}}
\newcommand{\ie}{{\em i.e.}}
\newcommand{\cm}{{{\cal M}}}
\newcommand{\cl}{{{\cal L}}}
\newcommand{\cd}{{{\cal D}}}
\newcommand{\cv}{{{\cal V}}}
\def\slashc{c\kern -.400em {/}}
\def\slashp{p\kern -.400em {/}}
\def\slashL{L\kern -.450em {/}}
\def\slashcl{\cl\kern -.600em {/}}
\def\Ww{{\mbox{\boldmath $W$}}}
\def\B{{\mbox{\boldmath $B$}}}
\def\noi{\noindent}
\def\nn{\noindent}
\def\sm{${\cal{S}} {\cal{M}}\;$}
\def\smn{${\cal{S}} {\cal{M}}$}
\def\nph{${\cal{N}} {\cal{P}}\;$}
\def\sb{$ {\cal{S}}  {\cal{B}}\;$}
\def\ssb{${\cal{S}} {\cal{S}}  {\cal{B}}\;$}
\def\ssbe{{\cal{S}} {\cal{S}}  {\cal{B}}}
\def\cviol{${\cal{C}}\;$}
\def\pviol{${\cal{P}}\;$}
\def\cpviol{${\cal{C}} {\cal{P}}\;$}

\newcommand{\lgg}{\lambda_1\lambda_2}
\newcommand{\lww}{\lambda_3\lambda_4}
\newcommand{\ppin}{ P^+_{12}}
\newcommand{\pmin}{ P^-_{12}}
\newcommand{\ppout}{ P^+_{34}}
\newcommand{\pmout}{ P^-_{34}}
\newcommand{\sinsq}{\sin^2\theta}
\newcommand{\cossq}{\cos^2\theta}
\newcommand{\yt}{y_\theta}
\newcommand{\hppll}{++;00}
\newcommand{\hpmll}{+-;00}
\newcommand{\hpplt}{++;\lambda_30}
\newcommand{\hpmlt}{+-;\lambda_30}
\newcommand{\hpptt}{++;\lambda_3\lambda_4}
\newcommand{\hpmtt}{+-;\lambda_3\lambda_4}
\newcommand{\dk}{\Delta\kappa}
\newcommand{\klam}{\Delta\kappa \lambda_\gamma }
\newcommand{\kac}{\Delta\kappa^2 }
\newcommand{\lac}{\lambda_\gamma^2 }
\def\gamgamtzz{$\gamma \gamma \ra ZZ \;$}
\def\gamgamtww{$\gamma \gamma \ra W^+ W^-\;$}
\def\gamgamtwwe{\gamma \gamma \ra W^+ W^-}

\def\stop{\tilde{t}}
\def\sto{\tilde{t}_1}
\def\stt{\tilde{t}_2}
\def\stl{\tilde{t}_L}
\def\str{\tilde{t}_R}
\def\msto{m_{\sto}}
\def\mstosq{m_{\sto}^2}
\def\mstt{m_{\stt}}
\def\msttsq{m_{\stt}^2}
\def\mt{m_t}
\def\mtsq{m_t^2}
\def\sint{\sin\theta_{\stop}}
\def\sintt{\sin 2\theta_{\stop}}
\def\cost{\cos\theta_{\stop}}
\def\sintsq{\sin^2\theta_{\stop}}
\def\costsq{\cos^2\theta_{\stop}}
\def\sinb{\sin\beta}
\def\cosb{\cos\beta}
\def\sinbb{\sin (2\beta)}
\def\cosbb{\cos (2 \beta)}
\def\tgb{\tan \beta}
\def\tgbsq{\tan^2 \beta}
\def\mqtt{\M_{\tilde{Q}_3}^2}
\def\mutt{\M_{\tilde{U}_{3R}}^2}
\def\sbottom{\tilde{b}}
\def\sbo{\tilde{b}_1}
\def\sbt{\tilde{b}_2}
\def\sbl{\tilde{b}_L}
\def\sbr{\tilde{b}_R}
\def\msbo{m_{\sbo}}
\def\msbosq{m_{\sbo}^2}
\def\msbt{m_{\sbt}}
\def\msbtsq{m_{\sbt}^2}
\def\mt{m_t}
\def\mtsq{m_t^2}
\def\selectron{\tilde{e}}
\def\seo{\tilde{e}_1}
\def\set{\tilde{e}_2}
\def\sel{\tilde{e}_L}
\def\ser{\tilde{e}_R}
\def\mseo{m_{\seo}}
\def\mseosq{m_{\seo}^2}
\def\mset{m_{\set}}
\def\msetsq{m_{\set}^2}
\def\me{m_e}
\def\mesq{m_e^2}
\def\snue{\tilde{\nu_e}}
\def\set{\tilde{e}_2}
\def\snul{\tilde{\nu}_L}
\def\msnue{m_{\snue}}
\def\msnuesq{m_{\snue}^2}
\def\smuon{\tilde{\mu}}
\def\smul{\tilde{\mu}_L}
\def\smur{\tilde{\mu}_R}
\def\msmul{m_{\smul}}
\def\msmulsq{m_{\smul}^2}
\def\msmur{m_{\smur}}
\def\msmursq{m_{\smur}^2}
\def\stau{\tilde{\tau}}
\def\stauo{\tilde{\tau}_1}
\def\staut{\tilde{\tau}_2}
\def\staul{\tilde{\tau}_L}
\def\staur{\tilde{\tau}_R}
\def\mstauo{m_{\stauo}}
\def\mstauosq{m_{\stauo}^2}
\def\mstaut{m_{\staut}}
\def\mstautsq{m_{\staut}^2}
\def\mtau{m_\tau}
\def\mtausq{m_\tau^2}
\def\gluino{\tilde{g}}
\def\mgluino{m_\tilde{g}}
\def\neuto{\tilde{\chi}_1^0}
\def\neutt{\tilde{\chi}_2^0}
\def\neutth{\tilde{\chi}_3^0}
\def\neutf{\tilde{\chi}_4^0}
\def\charop{\tilde{\chi}_1^+}
\def\chargtp{\tilde{\chi}_2^+}
\def\charom{\tilde{\chi}_1^-}
\def\chargtm{\tilde{\chi}_2^-}
\def\bino{\tilde{b}}
\def\wino{\tilde{w}}
\def\photino{\tilde{\gamma}}
\def\zino{tilde{z}}
\def\sdowno{\tilde{d}_1}
\def\sdownt{\tilde{d}_2}
\def\sdownl{\tilde{d}_L}
\def\sdownr{\tilde{d}_R}
\def\supo{\tilde{u}_1}
\def\supt{\tilde{u}_2}
\def\supl{\tilde{u}_L}
\def\supr{\tilde{u}_R}

\begin{titlepage}
\def\baselinestretch{1.2}
\topmargin     -0.25in

\vspace*{\fill}
\begin{center}
{\large {\bf SUSY Higgs at the LHC: Large stop mixing effects and
associated production }} \vspace*{0.5cm}

\begin{tabular}[t]{c}

{\bf G.~B\'elanger$^{1}$, F.~Boudjema$^{1}$,  and K.~Sridhar$^{2}$
}
 \\
\\
\\
{\it 1. Laboratoire de Physique Th\'eorique}
{\large LAPTH}
\footnote{URA 14-36 du CNRS, associ\'ee  \`a
l'Universit\'e de Savoie.}\\
 {\it Chemin de Bellevue, B.P. 110, F-74941 Annecy-le-Vieux,
Cedex, France.}\\

{\it 2. Department of Theoretical Physics, Tata Institute of
Fundamental Research} \\ {\it Homi Bhabha Road, 400 005 Mumbai,
India }\\
\end{tabular}
\end{center}

\centerline{ {\bf Abstract} }
\baselineskip=14pt
\noindent
 {\small We revisit the effect of the large stop mixing on the decay and
 production of the lightest SUSY Higgs at the LHC. We stress that whenever the inclusive
 2-photon signature is substantially reduced, associated production, $Wh$ and
 $t\bar t h$, with the subsequent decay of the Higgs into photons is
enhanced and becomes an
 even more important discovery channel. We also point out that these reductions in the
 inclusive channel do not occur for the smallest Higgs masses where the significance is
 known to be lowest. We show that in such scenarios the Higgs can be
 produced in the decay of the heaviest stop. For not too heavy masses of the pseudo-scalar
 Higgs where the inclusive channel is even further reduced, we show that large stop mixing
  also allows the production of the pseudo-scalar Higgs through stop decays.
These large mixing scenarios
 therefore offer much better prospects than previously thought. As a by-product we have
 recalculated $\sto \sto^* h$ production at the LHC and give a first evaluation of
 the $\sto \sto^* Z$. }
\vspace*{\fill}

\vspace*{0.1cm} \rightline{LAPTH-730/99} \rightline{TIFR/TH/99-17}
\rightline{{\large Apr. 1999}}

\end{titlepage}
\baselineskip=18pt

\setcounter{section}{1}

\setcounter{subsection}{0}
\setcounter{equation}{0}
\def\thesubsection {\thesection.\arabic{subsection}}
\def\theequation{\thesection.\arabic{equation}}

\setcounter{equation}{0}
\def\thequation{\thesection.\arabic{equation}}

\setcounter{section}{0} \setcounter{subsection}{0}

\section{Introduction}

The most popular alternative to the Standard Model, \smn, is
supersymmetry which at the moment fits in very well with all the
precision data. So well in fact that some see in the latest
precision data as preferring a low Higgs mass a very good evidence
for SUSY. In fact a low mass for one of the {\em scalar} Higgses
is the most robust limit of any supersymmetric model, contrary to
all the other (s)particles of the model which may have rather high
masses. In the minimal scenario of SUSY, the light Higgs mass can
not exceed $\sim 130$GeV. Considering the existing
LEP2\cite{Latest_mh_limit} direct searches which indicate a mass
greater than about $90$GeV means that the lightest SUSY Higgs is
confined to a small mass range. Yet this range of Higgs masses
poses considerable problems for hadron colliders. For a review
see\cite{Higgs_Hunter}. The dominant decay into $b \bar b$ is not
exploitable, especially in the {\em inclusive} production channel
$gg\ra h \ra b \bar b$, due to the huge QCD background. One
therefore has to rely on the much smaller two-photon
signal\cite{Hgaga_lhc}. However, especially for the LHC, the
two-photon decay of the light Higgs to which dedicated detectors
are being designed constitutes a challenge. Moreover many effects
either due to the
direct\cite{Htosusydecays,Baer_hgg_lhc,Baer_hsusymodes} or
indirect
(loop)\cite{Baer_hgg_lhc,Kileng_mixing,RggKane,AbdelStop_Hgg_Loops}
contributions of the rich SUSY spectrum enter the predictions of
the two-photon rate of the supersymmetric Higgs. These can lead to
a substantial reduction of the supersymmetric Higgs signal as
compared to the standard model Higgs. Take for instance the rather
simple scenario \cite{Baer_hgg_lhc,Class_H_Zwirner} where all
sparticles, apart from the parameters of the Higgs sector, are
very heavy and where mixing effects are negligible. This is the
scenario which has been extensively investigated by the
ATLAS\cite{ATLAS_hsusy,ATLAS_TDR}/CMS\cite{CMS_allHiggs}
collaborations which leads to the much celebrated $M_A-\tgb$ Higgs
discovery potential of the LHC. For short, we will refer to this
model as Class-H scenario. In this scenario the two-photon Higgs
signal can be much reduced compared to the \sm Higgs especially as
one lowers the mass of the pseudo-scalar boson, $A^0$.
Nonetheless, even in this scenario, this channel covers a large
part of the $M_A-\tgb$ discovery plane, while when $M_A$ gets
small so that the two-photon signal gets too small, one can extend
the discovery potential by exploiting the signatures of the then
not too heavy additional Higgses
\cite{Class_H_Zwirner,ATLAS_hsusy,ATLAS_TDR,CMS_allHiggs}. It is
therefore important to inquire how much  the important two-photon
signal can get reduced and equally important to investigate when
this reduction occurs, whether new mechanisms for Higgs production
open up or are enhanced. Could the latter then make up for the
loss in the former?


Considering that a general SUSY model furnishes an almost
untractable number of parameters to give an unambiguous answer,
apart from the Class-H scenario only partial
investigations\cite{Kileng_mixing,RggKane,AbdelStop_Hgg_Loops}
within specific models have been conducted. To quantify how the
rate of the two-photon signal can be affected as the SUSY
parameters are varied, it is instructive to take as a reference
point the signal for a \sm Higgs with a mass that of the lightest
SUSY Higgs. In\cite{RggKane} this has been done within the mSUGRA
hypothesis\cite{mSUGRA} but considering only the dominant
inclusive Higgs production channel: $gg \ra h \ra \gamma \gamma$.
One does find indeed that this ratio can be much smaller than
unity even for relatively large $M_A$ (which is generic in mSUGRA)
and hence making it more difficult to search for the SUSY Higgs
than for the same mass \sm Higgs. However it is
known\cite{Baer_htochi} that within mSUGRA other channels for
Higgs production may open up, like the cascading of the heavier
neutralino to a lighter one and a Higgs, thus offering the
fantastic possibility of not only discovering supersymmetry but
allowing an easy detection of the Higgs\cite{ATLAS_SUSYtoh} before
its observation in the two-photon channel. Recently, it has been
argued\cite{AbdelStop_Hgg_Loops}, that even in the large $M_A$
region, the so-called decoupling limit\cite{Haber_decoupling}, if
one introduces large mixing in the stop sector a very substantial
{\em reduction} can also ensue in the {\em inclusive} two-photon
Higgs signal. This effect together with the issue of the mixing in
the Higgsinos/gaugino sector had been studied previously by
comparing the rates with and without mixing\cite{Kileng_mixing}.
It was found that there were small regions in parameter space
where the rate for the two-photon Higgs signal could be either
very much {\em reduced} \underline{or} very much {\em enhanced} by
the inclusion of mixing.

When large reductions in an important channel occur it is crucial
to find out how other channels are affected. What has not been
stressed in the previous
studies\cite{Kileng_mixing,RggKane,AbdelStop_Hgg_Loops},
especially in the case of large mixing, is the importance of the
associated Higgs
production\cite{Associatedh_history,associatedhiggs_SM_early,associatedhiggs_SUSY_gaga_early}
and even if no efficient $b$-tagging were possible, how in these
scenarios these processes can salvage the Higgs signal. Within the
\sm and in the no-mixing scenarios, both
CMS\cite{Associated_h_CMS,CMS_allHiggs} and
ATLAS\cite{Wh_tth_ATLAS,ATLAS_TDR} have now shown that associated
Higgs production ($Wh, Zh$ and $t \bar t h$), with the subsequent
decay of the Higgs into photons, can provide an invaluable Higgs
signal, when enough luminosity has been accumulated. This is
because, although associated production has lower rates than the
inclusive channel, the corresponding signals are not plagued by as
much background. The CMS analysis\cite{CMS_allHiggs} for the \sm
Higgs shows that already with an integrated luminosity of $30
fb^{-1}$ the $\gamma \gamma l$ ($l=e,\mu$) leads to a significance
higher than $5$ (thus an observable Higgs signal) for the range of
light Higgs masses we are interested in. For a high luminosity of
$100 fb^{-1}$ this significance improves to more than $10$ and is
higher than the significance in the inclusive channel for
practically all Higgs masses in the range of interest. Although it
is known that the ATLAS analyses are less optimistic\footnote{The
differences between ATLAS and CMS are quantified in
\cite{Compar_CMS_ATLAS_hgaga}.} when it comes to the two-photon
signal, either in the associated or inclusive
channel\cite{ATLAS_TDR}, it remains that at high luminosity the
associated production provides a better reach in the $M_A-\tgb$
plane\cite{ATLAS_TDR}. One should therefore also inquire in the
case of the SUSY Higgs if the rates for associated production are
reduced {\em together} with the inclusive rates or if they can
rather help the discovery potential. At the same time if the rates
for the SUSY Higgs are very much affected this generally means
that some sort of non-decoupling of some of the SUSY particles is
taking place. These particles should then be observed directly.
Moreover since their coupling to the Higgs can not be negligible,
these same particles could trigger Higgs production, through their
decays for instance or through new associated productions. Another
important aspect to address is the impact of stop mixing on the
Higgs mass and its conjunction with the reduction in the inclusive
channel. Indeed, the significance in the inclusive two-photon
channel is very much dependent on the Higgs mass, even in the
narrow range allowed by SUSY\cite{ATLAS_TDR,CMS_allHiggs},
contrary to the associated two-photon channel where the
significances are rather flat as a function of the Higgs mass in
the range of interest\cite{CMS_allHiggs,ATLAS_TDR}.

The present paper revisits the case of the large mixing in the
stop sector\cite{Kileng_mixing,AbdelStop_Hgg_Loops}, how the
$Wh/Zh$ and $t\bar t h$ associated production saves the day when
the inclusive channel drops to critical levels and how  other new
channels for Higgs production open up. To set the stage, section 2
is intended as a reminder of how much a reduction in the usual
light Higgs signals can occur and is tolerable in the Class-H
scenario. This will serve as a benchmark when we study whether the
other scenarios could give reductions which are much worse than
those obtained with lowering $M_A$, a situation of some concern
especially if no new production mechanism is exploitable. We will
also present some approximations for evaluating the reductions due
to $M_A$ which will be useful even when we study the stop mixing
case.
\newline Our analysis of  the large mixing scenario in the stop
sector is contained in Section 3. We first consider the large
$M_A$ limit. While we confirm that large reductions in the
inclusive channel can occur, we point out that in most cases these
are no worse than what is obtained with a low $M_A$ in the
no-mixing case. Moreover we will show that if $\tgb>3$ an increase
in the {\em inclusive} channel is possible. This increase is not
possible for low values of $\tgb$ as studied in
\cite{AbdelStop_Hgg_Loops} because the effect is associated with a
too low Higgs mass already excluded by LEP2. We also carefully
analyse for which (light) Higgs masses these reductions occur. We
will show that contrary to the no-mixing scenario where the
signals in the inclusive channels are lowest for the lowest Higgs
masses, in the case of large stop mixing the most drastic drops in
the inclusive channels do not occur for the lightest Higgs mass
possible. Indeed the effect of mixing tends to increase the mass
of the Higgs compared to its value in the absence of mixing.
Considering that the significances for the \sm Higgs in the
inclusive channel are lowest for the lowest possible Higgs masses
in our range, $\sim 90-130$GeV, means that the largest reductions
do not necessarily correspond to the lowest Higgs signal. For
instance a reduction of $.4$ may well be tolerated for a Higgs
mass of $110$GeV but a reduction of $.8$ may be "too much" when it
occurs in conjunction with  $m_h=90$GeV. More importantly we find
that at the same time as the inclusive channel decreases, the
associated production increases and has much better significances
than with a \sm Higgs or with a corresponding SUSY Higgs where the
stop mixing have been switched off.  We will explain why this is
so. It should also be pointed out that the large reductions in the
inclusive channel occur mostly when one of the stops becomes
rather light, below about 200GeV. In many instances, as first
suggested by \cite{stophiggs_LHC}, associated $\sto \sto h$
production can provide a new channel to search for the light
Higgs. We will quantify how much one can benefit from this
additional channel. Most studies
\cite{Kileng_mixing,AbdelStop_Hgg_Loops} have assumed equality of
all soft squark masses which almost invariably leads to a maximal
mixing angle $|\sintt=1|$, where $\theta_{\stop}$ is the mixing
angle in the stop sector. Maximal mixing should be viewed as a
very special singular point in the large array of the SUSY
parameters and even though justified for the first two families as
suggested by the mSUGRA\cite{mSUGRA} scenario is quite unnatural
for the third family especially in view of the large Yukawa
coupling. We show, nonetheless, that maximal mixing is not always
required for the reductions in the two-photon rates to occur.
However, moving away even slightly from this singular mixing
angle, while not changing much the previous conclusions, can open
yet another Higgs production channel. We point out that provided
$\mstt$ is not too large, say $\mstt \leq 500$GeV, so that its
production rate is large, $\stt$ can provide a source of Higgs
through its decay into the lighter stop thanks to a sizeable
Yukawa $\stt \sto h$ coupling. This coupling is controlled by the
same parameters that make the $\sto \sto h$ coupling large and
which lead to a reduction of the inclusive channel. We will
compare the rate for this new Higgs production mechanism
$\sigma(pp \ra \stt \stt^* \ra \stt \sto^* h + \stt^* \sto h)$
with the associated lightest stop pair production mechanism
$\sigma(pp \ra \sto \sto^* h)$ \cite{stophiggs_LHC} and show that
the cascade decay of the $\stt$ can be substantial. This is
akin\cite{Baer_htochi} to the mixing in the higgsino-gaugino
sector which has been shown\cite{ATLAS_SUSYtoh} to allow a direct
Higgs production through the cascade decay $\chi^0_2 \ra \chi_1^0
h$~. We then move to the analysis of the combined effect of
allowing for smaller pseudo-scalar masses together with large stop
mixing. For moderate $M_A$ our conclusions are little changed, the
associated productions offering always a good channel. When $M_A$
gets rather small ($M_A \sim 250$GeV), the usual reduction, as
compared to the \smn, in both the inclusive and associated
production occurs. This is irrespective of mixing and can be
explained along the lines of what happens in the Class-H scenario.
Including the large mixing effects from the stops could decrease
even further the signal from the inclusive channel, but the same
effect again helps increase the associated production channel.
Therefore the reach in this channel alone is better than what
previously studied by ATLAS\cite{ATLAS_TDR,ATLAS_hsusy} and
CMS\cite{CMS_allHiggs} in the $M_A-\tgb$ plane for the no-mixing
Class-H scenario. Luckily in these situations with both a low
$M_A$ and large mixings in the stop sector we find that beside the
new channels for Higgs ($h$) production $\sigma(pp \ra \stt \stt
\ra \stt \sto h)$ and $\sigma(pp \ra \sto \sto^* h)$, one can also
have $\sigma(pp \ra \stt \stt \ra \stt \sto A)$. There are even
instances where the pseudo-scalar Higgs triggers $h$ production
through $A \ra Zh$. Independently of the extreme mixing scenario
studied here we advocate to exploit the potentially large Yukawa
coupling of the stops to search for the Higgs(es) through the
cascade decays of these third generation squarks. In all our
discussion we do not mention rescuing the Higgs signal through its
decay into $b \bar b$ in the associated
production\cite{tthtobb_LHC_tag} which would be possible provided
good $b$-tagging is available as discussed by
ATLAS\cite{ATLAS_htobb,ATLAS_TDR}. This is a difficult issue
\cite{CMS_htobb} especially at high luminosity and further
simulation studies are needed. Section 4 gives our conclusions.

\section{A warm up: Variation with $M_A$ in the case of no
mixing}

In order to compare the various effects of lowering the masses of
the SUSY particles, we start by briefly reviewing the situation
when the masses of all sparticles but those of the Higgs sector
are set to a high scale, $\tilde{M}_S=1TeV$. The mass of the
pseudo-scalar Higgs is let free. Moreover in the illustration we
have also taken the Higgs mixing parameter such that $\mu=-180GeV$
and the $SU(2)$ gaugino mass $M_2=500$GeV with the traditional GUT
assumption on the gaugino masses which at the electroweak
translates as
\beqn
M_1=\frac{5}{3} \tan^2 \theta_W M_2
\eeqn
Therefore strictly speaking we have allowed rather light charginos
and neutralinos. All the tri-linear $A$-terms were set to zero.
These kind of
scenarios\cite{Baer_hgg_lhc,Class_H_Zwirner,ATLAS_hsusy,CMS_allHiggs},
with high masses of sfermions, have been assumed in the simulation
searches for the Higgs(es) by the ATLAS/CMS Collaboration leading
to the much advertised $M_A-\tgb$ plots. Meanwhile it has been
known for some time that as $M_A$ increases one reaches a
decoupling regime\cite{Haber_decoupling} whereby at low energy
only the lightest neutral Higgs appears in the spectrum with the
important property that its couplings are essentially the same as
those of the standard model. This kind of \smn-like Higgs should
be easiest to discover at the LHC. However as the mass of the
pseudoscalar decreases the production rates of the lightest Higgs
also decrease. The reduction in the inclusive two-photon rate of
the Higgs, as compared to the \smn, is defined through the ratio

\beqn
\label{Rgg}
 R_{gg\gamma\gamma}=\frac{\Gamma^{SUSY}(h \ra g g) \times BR^{SUSY}(h
\ra \gamma \gamma)}{\Gamma^{SM}(h \ra g g) \times BR^{SM}(h \ra
\gamma \gamma)}
\eeqn

\begin{figure*}[htbp]
\begin{center}
\mbox{\epsfxsize=14cm\epsfysize=8cm\epsffile{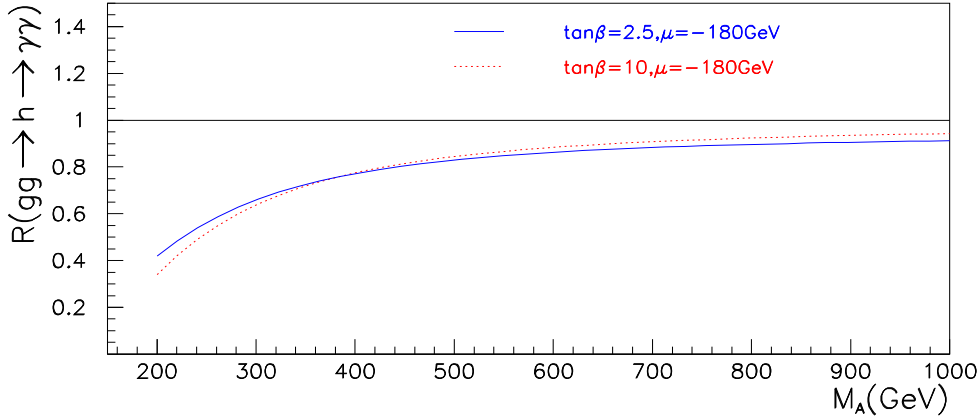}}
\vspace*{-1cm} \caption{\label{matgb2.5_MA_Rgg}{\em Variation of
$R_{gg\gamma\gamma}$ with $M_A$, for $\tgb=2.5$ (full) and
$\tgb=10$ (dotted)}} 
\end{center}
\end{figure*}

\begin{figure*}[htb]
\begin{center}
\mbox{\epsfxsize=16cm\epsfysize=14cm\epsffile{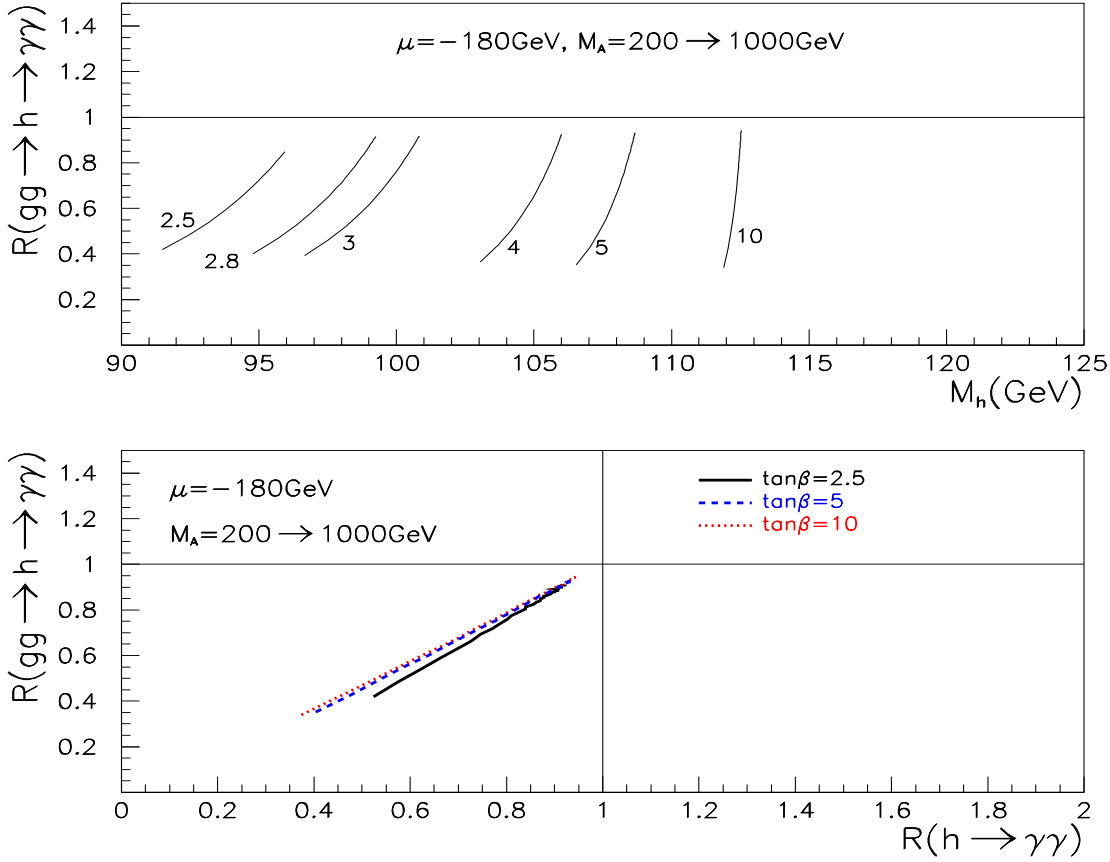}}
 \caption{\label{matgb2.5_Mh_Rgaga}{\em
a) $R_{gg\gamma\gamma}$ {\it vs} $M_h$ as $M_A$ varies from 200GeV
to 1000GeV for different values of $\tgb$. $\mu=-180$GeV. The
lowest values of $R_{gg\gamma\gamma}$ and $R_{\gamma\gamma}$
correspond to $M_A=200$GeV. b) $R_{gg\gamma\gamma}$ {\it vs}
$R_{\gamma\gamma}$}.}
\end{center}
\end{figure*}

This ratio is calculated by taking the {\it same mass} for the \sm
Higgs as the one that is derived for the SUSY Higgs once all the
SUSY parameters are set. Throughout this paper we use {\tt
HDECAY}\cite{HDECAY} to calculate all the couplings, widths and
branching ratios of the Higgs. This program incorporates the
leading two-loop corrections for the Higgs masses
following\cite{CarenaWagner_Higgs_Approx1}. We show in
Fig.~\ref{matgb2.5_MA_Rgg} how this ratio decreases with $M_A$.
This ratio can drop to as little as $\sim 30\%$ for $M_A=200$GeV
and $\tgb=10$. Though trivial in this case, it is useful to point
for later that as the pseudo-scalar Higgs mass decreases so do the
other Higgs masses\footnote{In the analysis we have required
$M_h>90$GeV.}, therefore the most drastic drops occur for the
lowest range of the lightest Higgs, see
Fig.~\ref{matgb2.5_Mh_Rgaga}. This is particularly drastic for
$\tgb=2.5$GeV, where the drop occurs around $m_h \sim 90$GeV. It
is for these low masses that the significance of the \sm Higgs is
also lowest\cite{ATLAS_hsusy,CMS_allHiggs} and therefore for this
low $\tgb$ this would constitute the worst scenario for the
discovery of the lightest SUSY Higgs through its two-photon
decay\footnote{Of course, for $M_A \leq 2 m_t$ there is a chance
of discovering the other Higgses.}.

What is troublesome for a low $M_A$ is that the branching fraction
into two photons is the main reason behind the drop, as shown in
Fig.~\ref{matgb2.5_Mh_Rgaga}. This ratio is defined as

\beqn
\label{Rgamgam}
 R_{\gamma\gamma}=\frac{ BR^{SUSY}(h
\ra \gamma \gamma)}{BR^{SM}(h \ra \gamma \gamma)}
\eeqn

\noi For instance for $M_A=200$GeV and $\tgb=2.5$, the ratio of the
branching fraction into photons, $R_{\gamma\gamma}$, is reduced to
about .5 with respect to what it would be in the \sm. This
reduction accounts for much of the reduction in $R_{gg\gamma
\gamma}$, $R_{gg\gamma \gamma}=.4$. Therefore one expects also a
considerable drop in the Higgs signal even in the associated
channel $Wh$ and $t \bar{t} h$ with the subsequent decay of the
Higgs into two photons. These channels have been shown to be
invaluable\cite{Wh_tth_ATLAS,ATLAS_hsusy,ATLAS_TDR,CMS_allHiggs}
especially when a high luminosity has been accumulated.
%

To get an understanding of these gross features and compare with
what happens in other scenarios, it is worth discussing how the
various couplings, $t \bar{t} h$, $b \bar b h$ and $Wh/Zh$ that
enter both the associated production and, at the loop level, the
inclusive production are influenced by a change in $M_A$. This is
best illustrated and most transparent in the large $M_A$,
so-called decoupling, limit which has been shown to be already
operative at $200$GeV\cite{Haber_decoupling}. Take the $t \bar t
h$ coupling which differs from the \sm by the factor $R$:

\beqn
V_{t t h}=\frac{g}{2 M_W} R \; \mt \;\;\;\; {\rm with}\;\;\;\;
R=\frac{\cos \alpha}{\sin \beta}
\eeqn
where $\alpha$ is the usual angle that appears in the
diagonalization of the CP-even neutral Higgs mass matrix. As was
shown elsewhere\cite{eenous_t1t1h,Wells_Higgs98}, in this limit
and {\em up to radiative corrections} we may introduce the factor
$r$

\beqn
\tan \alpha \tan \beta =- (1+r) \;\;\;{\rm with} \;\;\; r \ll 1
\eeqn

\noi where $r$ collects all $M_A$ dependence and other radiative corrections which also
occur in the computation of the Higgs masses. Neglecting the
latter we have

\beq \label{rdecoupling}
 r \simeq \frac{2 M_Z^2}{M_A^2} \;\frac{\tgbsq -1}{\tgbsq+1} \ge 0
\eeq

then the {\it reduction} factor which appears in $t \bar{t} h$ is
\beqn
R^2=\frac{1+\tan^2 \beta}{1+\tan^2 \beta + r^2 +2 r}\;\;\;\; R
\simeq 1-\frac{r}{1+\tan^2 \beta}
\eeqn

Likewise it is found that in $h \bar{b} b$ there is an {\em
enhancement} factor which especially, for larger values of $\tgb$,
is more substantial than the reduction in the top vertex
\beq
\label{bbh}
R_{bbh}\simeq 1+r \frac{\tgbsq}{1+\tgbsq}
\eeq

On the other hand the $WWh/ZZh$ vertex, controlled by $\sin(\alpha
-\beta)$,  is much less affected: it only shows a {\em quadratic}
dependence in $r$:
\beqn
\label{Zzh}
R_{VVh}\simeq 1\;-\;\frac{r^2}{2} \frac{\tgbsq}{(1+\tgbsq)^2}
\eeqn


In the \sm $\Gamma(h \ra \gamma \gamma)$ is dominated by the $W$
loop which interferes destructively with the top. Since, in this
scenario the dominant $W$ coupling is hardly affected at moderate
$M_A$ the little change in the top (bottom) coupling has
negligible effect on $\Gamma(h \ra \gamma \gamma)$. However this
is not the case for the branching fraction into photons. Here,
since the total width is dominated by the width into $b \bar b$,
which is larger than in the \smn, the branching ratio into photons
will be reduced, especially as $\tgb$ increases, see
Eq.~\ref{bbh}. On the other hand we expect a slight decrease in
the $\Gamma(h \ra gg)$. This is because it is dominated by the top
loop in the \smn, and therefore it is reduced roughly as the
$t\bar t h$ vertex is reduced. Therefore the main effect in the
production rate $pp \ra h \ra \gamma \gamma$ is due to the
reduction in $Br(h \ra \gamma \gamma)$. This very crude argument
gives the correct order of magnitude in the different drops in
$R_{gg\gamma\gamma}$ and $R_{\gamma\gamma}$ shown in the figures.
Writing for example
\beqn
\label{RggmA}
R_{\gamma \gamma} \simeq 1\;-\;\frac{\Gamma^{SM}(h\ra b \bar
b)}{\Gamma^{SUSY}(h\ra b \bar b)} \simeq 1-\frac{4 M_Z^2}{M_A^2}
\frac{\tgbsq (\tgbsq-1)}{(1+\tgbsq)^2}
\eeqn
we recover $R_{\gamma \gamma}=.483$ for $M_A=200$GeV and
$\tgb=2.5$ which compares very well with the full calculation.
Moreover in a first approximation, the change in the width into
gluons can be mostly accounted for by the change in the $t\bar t
h$ vertex. In which case we may write
\beqn
\label{RglgluggmA}
R_{g g \gamma \gamma}\simeq 1-\frac{4 M_Z^2}{M_A^2} \frac{
\tgbsq-1}{1+\tgbsq}
\eeqn

For larger $\tgb$ and especially for low values of $M_A$ the
approximation is acceptable but not as good. This is partly due to
the effect of radiative corrections on the $h \ra b \bar b$
coupling through the diagonalisation of the neutral Higgs mass
matrices\footnote{In this discussion this applies especially to
the off-diagonal terms of the Higgs mass matrix.}. Especially for
large $\tgb$ these corrections are no longer so suppressed
compared to the $M_A^2$ corrections
\cite{RggKane,Baer_Wells,Wells_Higgs98,CarenaWagner_Tevatron}. In
our case the effect is rather marginal since the only mixing
parameter, $\mu$, is rather small compared to the SUSY scale.
However  let us stress that in all the analyses in this paper even
when considering large values of the tri-linear coupling (see next
section) the branching ratio into $b \bar b$ is hardly affected.
Because our aim is to concentrate on the effect of the tri-linear
coupling of the top sector we do not, in the present paper,
analyse the case with very large $\tgb$ as these would require to
analyse the sbottom sector and also for large $\mu$ possible
reductions in the $h \ra b \bar b$ branching ratio.

\begin{figure*}[hbtp]
\begin{center}
\mbox{\epsfxsize=16cm\epsfysize=16cm\epsffile{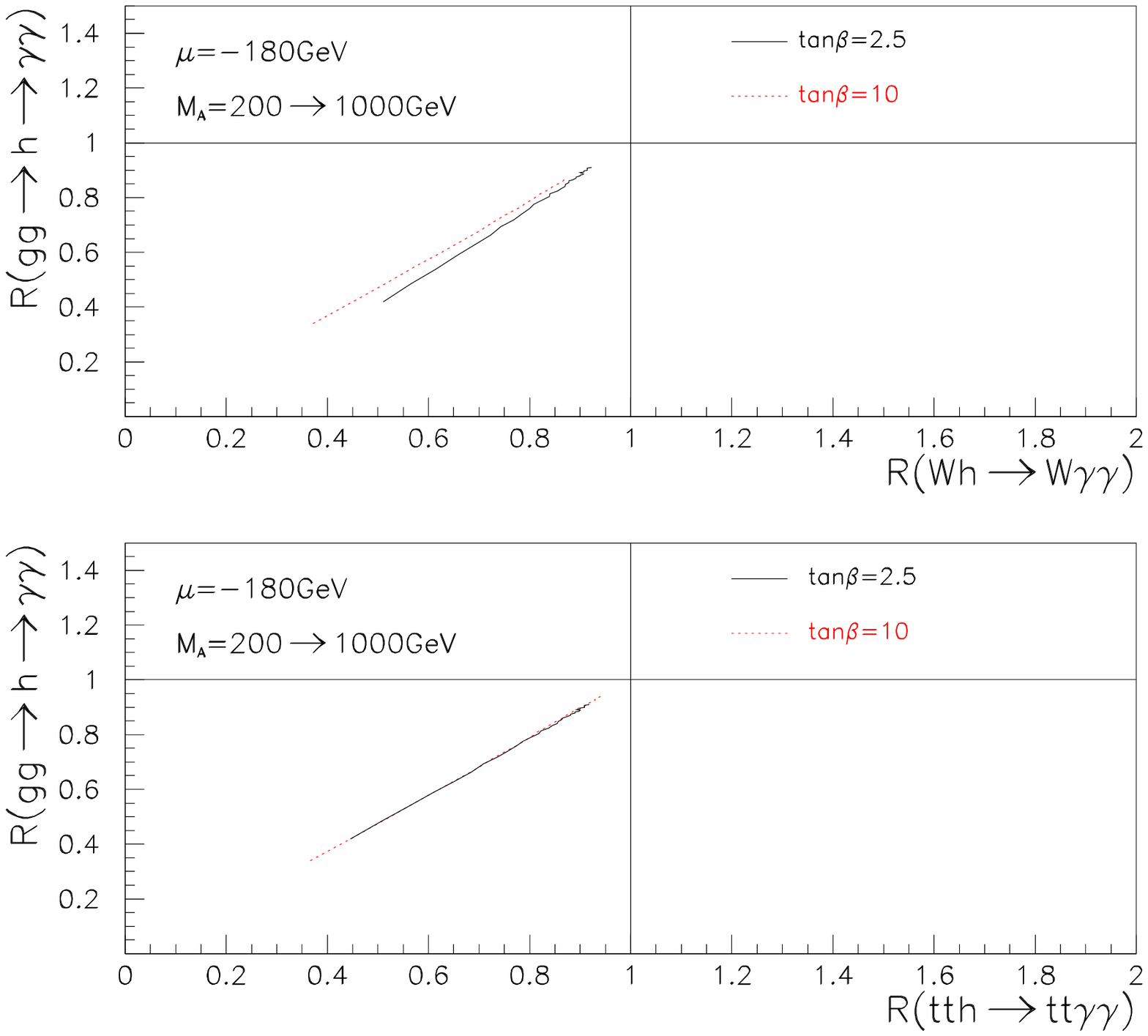}}
\caption{\label{matgb2.5_wwh_Rgg}{\em Variation of
$R_{gg\gamma\gamma}$ {\it vs} $R_{W\gamma\gamma}$-$R_{t\bar{t}
\gamma\gamma}$ with $200\le M_A \le 1000$GeV and $\tgb=2.5
(full),10 (dotted)$.}}
\end{center}
\end{figure*}

When considering the associated channels, beside the reduction in
the two-photon branching ratio, a further, even though slight,
reduction factors affects $t \bar t h$ production while we expect
$Wh$ to be much less affected. This is borne out by the numerical
analysis shown in
Fig.~\ref{matgb2.5_MA_Rgg}-\ref{matgb2.5_wwh_Rgg}. Once again we
define, for the associated productions, ratios normalized to the
\sm rates for the same Higgs mass:
\beqn
\label{whgaga} R_{W\gamma\gamma}=\frac{\sigma^{SUSY}(pp \ra Wh)
\times BR^{SUSY}(h \ra \gamma \gamma)}{\sigma^{SM}(pp \ra Wh)
\times BR^{SM}(h \ra \gamma \gamma)}
\eeqn
and similarly for the associated top production:
$R_{tt\gamma\gamma}$. At the level of the cross sections,
$\sigma(pp \ra Wh, Zh, t\bar t h)$, the ratios are assumed to be
given by the ratios of the squares of the $W W h$ and $t \bar t h$
couplings. We clearly see,
Figs.~\ref{matgb2.5_MA_Rgg}-\ref{matgb2.5_wwh_Rgg}, that a
lowering of $M_A$ in case of no mixing not only results in a
lowering of $m_h$ but also in a reduction of {\em both} the
inclusive and associated two photon channels, as compared to the
\sm signal for the same $m_h$. The worst hit channels are the
direct production and the $t\bar th$. The $Wh$ channel is slightly
less affected. Note that $t \bar t\gamma\gamma$ {\it vs}
$R_{gg\gamma \gamma}$ shows almost no $\tgb$ dependence,
Fig.~\ref{matgb2.5_wwh_Rgg}. This is due to the dominance of the
top loop in the $gg \ra h$ production, controlled by the same
vertex that enters the associated $t \bar t h$ cross section.
These reductions occur for the lightest Higgs mass and are due
essentially to the drop in the branching ratio of the Higgs into
two photons, see
Figs.~\ref{matgb2.5_MA_Rgg}-\ref{matgb2.5_wwh_Rgg}. Since as we
pointed earlier the significances in the associated channels are
rather flat with respect to the Higgs mass in the range we are
interested in, this explains why the $5\sigma$ discovery region
based on the associated channel in the $M_A-\tgb$
plane\cite{ATLAS_hsusy,CMS_allHiggs} are almost independent of
$\tgb$ ($2.5 \leq \tgb \leq 10$). On the other hand the discovery
region based on the inclusive channel shows a strong difference
between low and high $\tgb$ values. This is due essentially to the
low significances for low Higgs masses which translates into low
significances for $\tgb$ in case of no mixing for a fixed value of
$M_A$. Therefore although the {\em reduction} due to a low $M_A$
is slightly worse for $\tgb=10$ than for $\tgb=2.5$,
Fig.~\ref{matgb2.5_MA_Rgg}, the significance in the direct channel
is much better for $\tgb=10$ ($m_h \sim 110$GeV) for $\tgb=2.5$
($m_h \sim 93$GeV). This observation is to be kept in mind and
shows the importance of localising where in terms of $m_h$ any
reduction, especially, in the inclusive direct channels occurs.
Take for instance the CMS analysis\cite{CMS_allHiggs}. It is found
that already with a low luminosity of $30$fb$^{-1}$ the
significance for the \sm Higgs is larger in the associated channel
than in the inclusive channel for $m_h<105$GeV and is above $5$.
Translated to Class-H this means that for $M_A \geq 450$GeV
associated production allows observability of the light Higgs for
all values $\tgb=2.5-10$ whereas direct production extends the
reach in $M_A$ for $\tgb=10$ ($M_A \sim 400$GeV). With a higher
luminosity of $100$fb$^{-1}$, the CMS analysis shows that the
reach in $M_A$ is better in the associated channel for all values
of $\tgb$ and especially for low $\tgb$. In terms of the ratio
$R_{\gamma \gamma}$, this analysis translates into discovery for
$R_{\gamma \gamma} > .4$ corresponding to $M_A > 220$GeV (even
slightly better for $\tgb=2.5$, see Fig.~\ref{matgb2.5_wwh_Rgg}).
Note that one can recover the obsvervability region of the SUSY
Higgs of the CMS analysis by combining our results for the ratios
$R$ with their analysis for the \smn. As stated earlier the
ATLAS\cite{ATLAS_TDR} analysis requires higher luminosities and
the above numbers correspond roughly to a luminosity of
$300$fb$^{-1}$ to take full advantage of the associated
production\footnote{For lower luminosities the ATLAS significances
in the associated channels are based on a Poisson statistics. We
thank Guillaume Eynard for providing us with his code and the
"data" for the SM Higgs in the separate channels $Wh/Zh$ and $t
\bar t h$, see also \cite{Wh_tth_ATLAS}.}.

\section{Mixing in the stop sector}
 To discuss the stop sector and define our conventions, we turn to the weak
eigenstate basis where the mass matrix in the $\stop_L,\stop_R$
involves the SUSY soft-breaking masses: the common $SU(2)$ mass
$\tilde{m}_{\tilde{Q}_3}$ and the $U(1)$ mass
$\tilde{m}_{\tilde{U}_{3R}}$, beside the mixing,
$\tilde{m}_{\tilde{t}_{LR}}^2$

\beqn
m_{\tilde{t}_L}^2&=&\tilde{m}_{\tilde{Q}_3}^2 + m_t^2 +
\frac{1}{2} M_Z^2  (1- \frac{4}{3} \sww) \cos(2 \beta)
\\ m_{\tilde{t}_R}^2&=&\tilde{m}_{\tilde{U}_{3R}}^2 + m_t^2 +
\frac{2}{3} M_Z^2 \sin^2 \theta_W \cosbb \nonumber
\\ m_{\tilde{t}_{LR}}^2&=&-m_t (A_t
+\frac{\mu}{\tgb}) \equiv -m_t \tilde{A}_t
\eeqn

One sees that apart  from the soft SUSY-breaking parameters:
$\tilde{m}_{\tilde{Q}_3}$, $\tilde{m}_{\tilde{U}_{3R}}$ and the
tri-linear top term ($A_t$), there appears also the ubiquitous
$\tgb$ and the higgsino mass term $\mu$.

The stop mass eigenstates are defined through the mixing angle
$\theta_{\tilde{t}}$, with  the lightest stop, $\sto$,
\beq
\sto=\cost \; \stl + \sint \; \str
\eeq
It is quite useful to express  the mixing angle
as\cite{eestopstop_polar_Nojiri,Stops_Porod}:

\beqn
\label{s2t}
 \tan (2 \theta_{\stop})= \frac{-2 m_t \tilde{A}_t}{\tilde{m}_{\tilde{Q}_3}^2-
\tilde{m}_{\tilde{U}_{3R}}^2+\frac{M_Z^2 \cos2\beta}{2}(1-\frac{8
s_W^2}{3})}\;\;\;\;\; {\rm or} \;\;\;\;\sin (2 \theta_{\stop})=
\frac{2 \; m_{\tilde{t}_{LR}}^2}{\mstosq-\msttsq}
\eeqn

For further reference note, in the case of equal soft SUSY
breaking masses for the left and right sector of the stop (
$\tilde{m}_{\tilde{Q}_3}^2=\tilde{m}_{\tilde{U}_{3R}}^2$), that
apart from the case of extremely small mixing $\tilde{A}_t=
{\cal{O}}(M_Z/10)$, one has maximal mixing: $\sin(2
\theta_{\stop})\simeq 1$. In this case we have

\beqn \tan (2 \theta_{\stop})\simeq \frac{m_t}{ M_Z}\frac{12 \tilde{A}_t}{M_Z}
\frac{\tgb^2+1}{\tgb^2-1}\eeqn

\subsection{The $\sto \sto h$ vertex}
Mixing in the stop sector not only allows one of the stops to be
rather light, but this light stop can have rather large Yukawa
couplings. Let us therefore discuss this coupling.
 The stop-stop Higgs couplings, like the stop
mass matrix, emerge essentially from the F-terms in the scalar
potential (there is a residual $D$ term component $\propto \mzz$).
With the angle $\alpha$ in the Higgs mixing matrix, the $\sto \sto
h$ coupling is (we write the potential)

\beqn
\label{stopstophcoupling} V_{\sto \sto h}&= & -g
\frac{\mt}{\mw} \frac{\cos \alpha}{\sin \beta} \bigg( (A_t-\mu
\tan \alpha) \sint \; \cost \;-\; \mt \nonumber \\ &+&
\frac{\mzz}{\mt} \frac{\sin \beta}{\cos \alpha} \sin(\alpha+\beta)
\left((\frac{1}{2}-\frac{2}{3} \sww)\costsq +\frac{2}{3} \sww
\sintsq \right)  \biggr)\nonumber \\
\eeqn

The vertex does involve some important parameters which stem from
the Higgs sector, notably the angle $\alpha$. In the decoupling
limit \cite{Haber_decoupling} which we are most interested in and
up to radiative corrections Eq.~\ref{stopstophcoupling} writes

\beqn
\label{stopstophcoupling2} V_{\sto \sto h}&=& +g R \frac{1}{\mw}
\biggl\{  \mt^2 +  \sint \; \cost \left( \sint \; \cost
(\mstosq-\msttsq) -\frac{\mt \;\mu \;r}{\tgb} \right) \nonumber \\
 &+& \mzz ((2+r)\cos^2\beta - 1)
\left((\frac{1}{2}-\frac{2}{3} \sww)\costsq +\frac{2}{3} \sww
\sintsq \right) \biggr\} \nonumber \\
\eeqn

\noindent We see that in the limit $r \ll 1$ where $r$ is neglected, the $\sto \sto h$
very much simplifies. Note that neglecting the correction due to
$r$, the coupling no longer depends on $\mu$. Notice also that
Eq.~\ref{stopstophcoupling2} shows that
 this
correction is reduced as $\tgb$ gets larger. Discarding the $r$
correction altogether, we end up with a compact formula

\beqn \label{approxtth}
 V_{\sto \sto h} &\simeq &\frac{g}{\mw} \biggl( \sin^2(2 \theta_{\stop})
 \frac{(\mstosq-\msttsq)}{4} \;+\; \mt^2 \nonumber \\
 &+& \mzz \cos (2
\beta)\left((\frac{1}{2}-\frac{2}{3} \sww)\costsq +\frac{2}{3}
\sww \sintsq \right) \biggr)
\eeqn

 We also
confirm that the $\tgb$ dependence in the vertex is also hardly
noticeable. Eq.~\ref{approxtth} makes it clear that even for
maximal mixing, $\sin^2 2\theta_{\stop} \sim 1$ the contribution
of the stops and that of the top cancel each other thus leading to
a very small vertex.  The dip occurs for values of the mixing
angle such that:

\beq
\label{dipinvertex}
\sin^2 2\theta_{\stop}\simeq\frac{4 \mt^2}{\mstt^2-\msto^2}
\eeq
On the other hand when the mixing is negligible, the vertex is
accounted for almost entirely by the top mass and therefore has
the same strength as the $t t h$ vertex.

The $\stt \stt h$ vertex can be obtained from $\sto \sto h$ by
$\sin \theta_{\stop} \leftrightarrow \cos \theta_{\stop}$ and
$\msto \leftrightarrow \mstt$. Therefore if the $\stt \stt h$ and
$\sto \sto h$ vertices were to be added, the mixing terms do not
survive, as expected since the latter mix the left and right
states. This is to be kept in mind. In situations where the stop
masses are of the order of the top mass so that they both
contribute to $h\ra gg$ or $h\ra \gamma \gamma$, the effect of
mixing will, to a large extent, be washed away.

Already at this point we can attempt to predict the general
features in $R_{gg\gamma \gamma}$ and $R_{\gamma \gamma}$ that
will be introduced by large mixing in the stop sector. Consider
the large $M_A$ limit where the $\sto \sto h$ vertex is most
transparent, see Eq.~\ref{approxtth}. Naturally the stop will
contribute if its mass is not too large and if its coupling to the
Higgs is also large. When there is no mixing, only the diagonal
$m_t^2$ term in Eq.~\ref{approxtth} will, in both $\Gamma(h \ra
gg)$ and $\Gamma(h\ra \gamma \gamma)$, interfere {\it
constructively} with the top quark contribution. We therefore
expect an enhancement of $\Gamma(h \ra gg)$, that is of the
inclusive production. On the other hand, the fact that the
top/stop loops and $W$ interfere destructively, means that
$\Gamma(h\ra \gamma \gamma)$ will get smaller. Nonetheless since
the $W$ loop is much larger than the top loop, the reduction in
the two-photon decay width will be modest compared to the
enhancement in the two gluon width. Considering that at large
$M_A$ the width into $b\bar b$ (thus the total width) is hardly
affected by mixing and hence sensibly the same as in the \smn,
direct production $\sigma(pp \ra h \ra \gamma \gamma)$ is
enhanced. At the same time associated $Wh/Zh$ and $t\bar t h$ with
the subsequent two-photon decay of the Higgs will be reduced
somehow. For moderate mixing the $\sto \sto h$ vertex gets
vanishingly small: here no effect is to be expected, either in any
of the associated productions nor in the direct production. When
the mixing gets very large so that now, it is the term in
$\mstt^2$ in Eq.~\ref{approxtth} which dominates, the sign of the
interferences between the stop and the top quark loop gets
reversed. In this situation direct production can get extremely
small, the stop loop cancelling the top loop. In the two photon
decay, on the other hand when this cancellation takes place it
still leaves the large $W$ contribution. Nonetheless, the increase
in $R_{\gamma\gamma}$ will be modest compared to the dramatic
decrease in $R_{gg\gamma\gamma}$. Since the total width is hardly
affected by these mixing effects the direct inclusive production
will be much reduced. However associated $Wh/Zh$ and $t\bar t h$
gets enhanced in these situations.

\subsection{$pp \ra \sto \sto^* h$ at the LHC}

\begin{figure*}[p]
\begin{center}
\mbox{\epsfxsize=16cm\epsfysize=18cm\epsffile{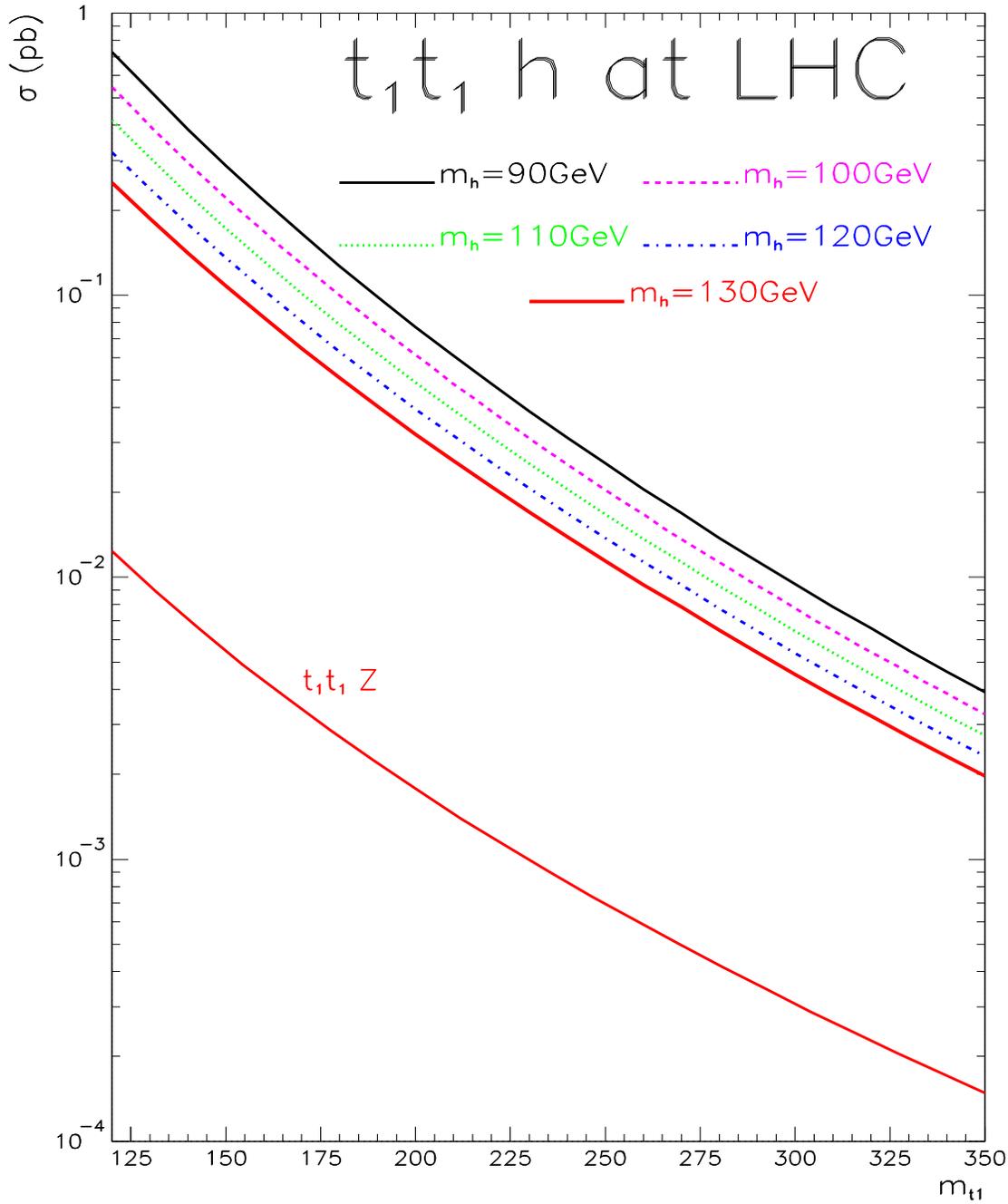}}
\caption{\label{t1t1h_mt1}{\em $\sto \sto h$ at the LHC as a
function of $\msto$ and for a range of SUSY Higgs masses. The
$\sto \sto h$ vertex is set in the limit of large $M_A$ with no
mixing and no D-term, see text of how to normalise it when the
SUSY parameters are fixed. Also shown is $\sto \sto Z$. For the
latter the vertex has been computed with $\cos^2
\theta_{\stop}=1/2$, {\it i.e.} maximal mixing. For other values
of the mixing, rescale by using the vertex $(\cos^2
\theta_{\stop}/2 -2/3 s_W^2)$. We have taken the CTEQ4 structure
function with a scale set at the invariant mass of the
subprocess.}}
\end{center}
\end{figure*}

\begin{figure*}[p]
\begin{center}
\mbox{\epsfxsize=16cm\epsfysize=18cm\epsffile{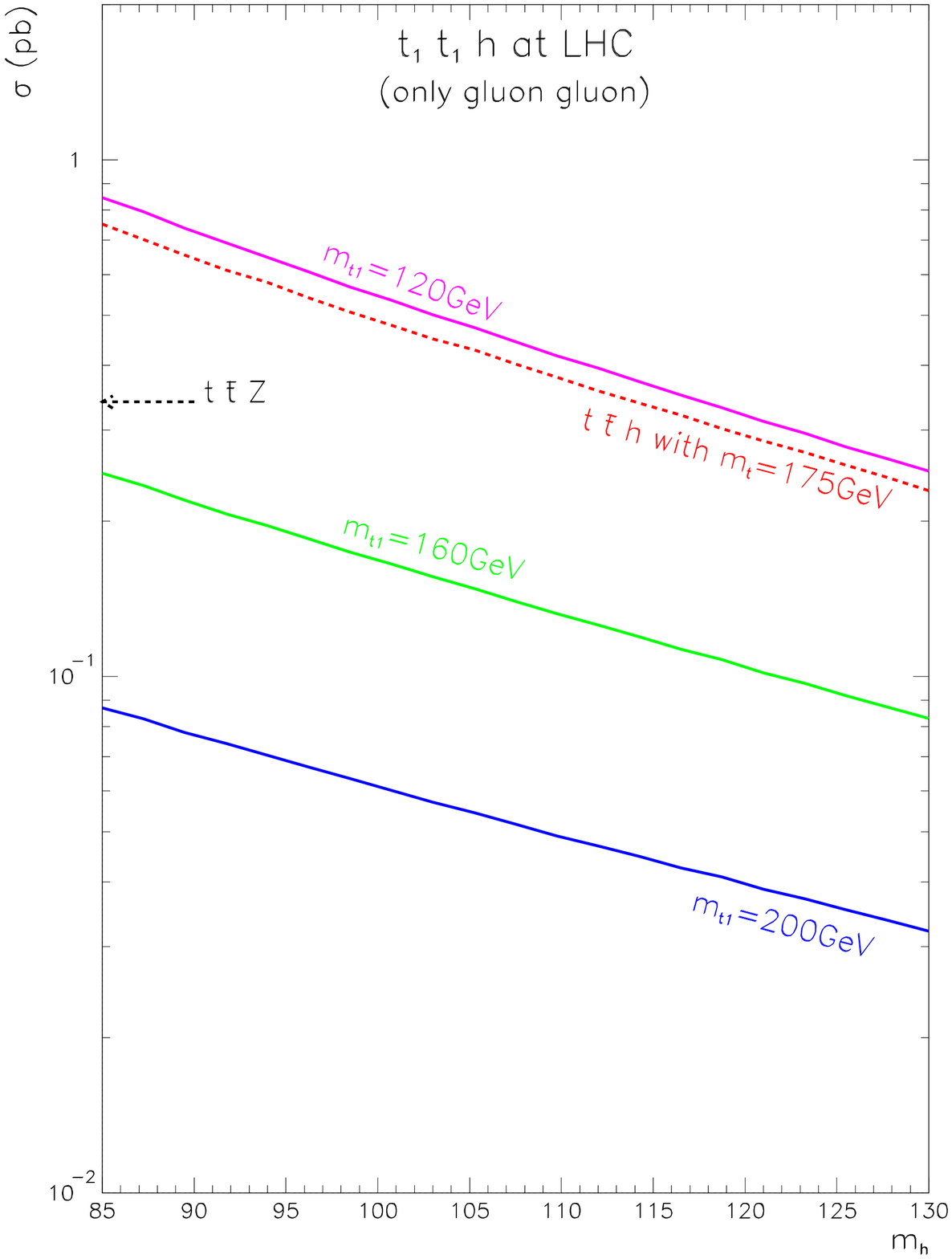}}
\caption{\label{t1t1h_mh}{\em As in Figure~\ref{t1t1h_mt1} but as
function of the Higgs mass. The $t \bar t h$ is also shown for
comparison for the same set of structure functions and by only
taking into account the gluon gluon processes. For the latter
including the small quark initiated process, our results agree
with \cite{Higgs_xsections_update}. Also shown is $t \bar t Z$.}}
\end{center}
\end{figure*}

Because $\sto$ is relatively light and its coupling to the Higgs
(h) large, associated stop cross sections can, exceptionally, be
of the order of that of the associated top cross
section\cite{stophiggs_LHC} or even larger. At the LHC this cross
section is essentially induced by gluon gluon fusion and is
therefore directly proportional to the square of the $\sto \sto h$
vertex. We have recalculated this cross section with the help of a
modified version of {\tt CompHep}\cite{comphep} to properly take
into account the radiative corrections to the Higgs mass and
couplings. For our analysis we have found it useful to calculate
the cross section at the LHC by taking, as a reference point, the
$m_t^2$ term only in the $\sto \sto h$ vertex,
Eq.~\ref{approxtth}. The cross section can then be easily
evaluated by specifying as independent input parameters $\msto$
and $m_h$ only. The corresponding cross sections are shown in
Fig.~\ref{t1t1h_mt1} and Fig.~\ref{t1t1h_mh}. We have made a
polynomial fit, in the variables $m_h-\msto$ to these cross
sections that reproduces the full results with a precision better
than $2\%$, which is  well within the uncertainty due to the
choice of scale and structure function.  Once a set of SUSY
parameters is given, apart from the stop masses and $\tgb$ it will
also furnish the corresponding Higgs mass, $m_h$, and the proper
$\sto \sto h$ vertex can be evaluated. One can then properly
normalise our cross sections. Considering the relative complexity
of the $pp \ra \sto \sto h$ cross section this method is much more
efficient when we are scanning over many SUSY parameters as done
in the present analysis since we do not have to recalculate the
$pp \ra \sto \sto h$ for each scan. Our results agree with those
shown in \cite{stophiggs_LHC} as well as in
\cite{Moretti_stophiggs}, however the largest cross sections shown
in \cite{stophiggs_LHC} do not pass our constraint on the Higgs
mass $m_h>90$GeV and/or $\Delta\rho$ (see below).

\subsection{The $\stt \sto h,A,H$ vertex and $\stt \ra \sto h,H,A$}
The $\sto \stt h$ vertex may be cast into
\beqn
\label{stop1stop2hcoupling2} V_{\sto \stt h}&=& +g R
\frac{1}{\mw} \biggl\{ \frac{\cos 2\theta_{\stop}}{4} \left(\sin
2\theta_{\stop}\; (\mstosq-\msttsq) -\frac{2 \mt \;\mu \;r}{\tgb}
\right) \nonumber \\ &+& \mzz \sin 2\theta_{\stop}\; (\cos 2\beta
+ r \cos^2 \beta)\left(\frac{2}{3} \sww-\frac{1}{4} \right)
\biggr\} \nonumber \\ &\ra&  +g R \frac{1}{4 \mw} \sin 4
\theta_{\stop} \; (\mstosq-\msttsq)
\eeqn

\begin{figure*}[htbp]
\begin{center}
\mbox{\epsfxsize=14cm\epsfysize=15cm\epsffile{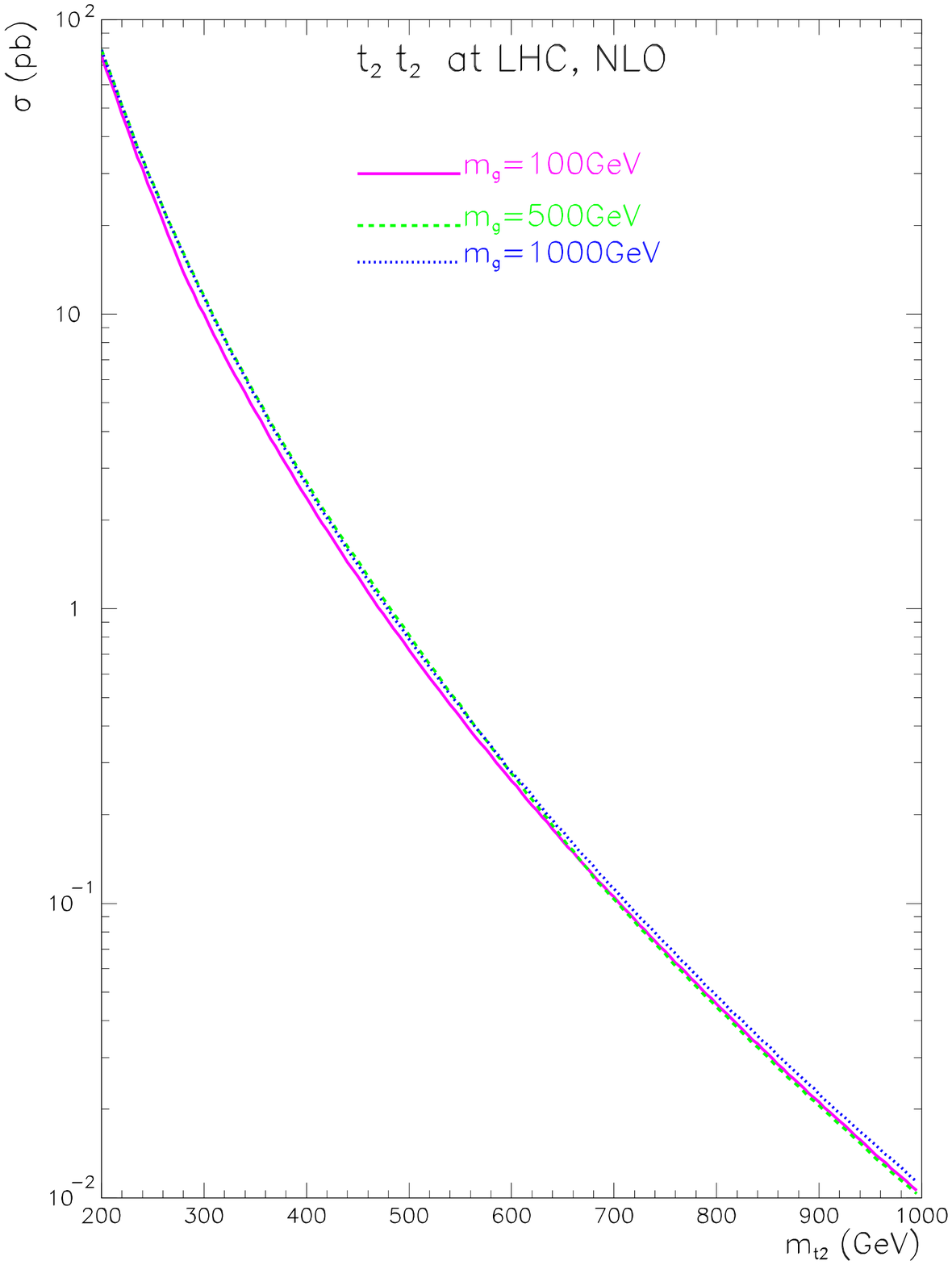}}
\caption{\label{t2t2_lhc}{\em Next-to-leading order $\sto$ pair
production at the LHC, for three representative values of the
gluino mass. We used the code given to us by Michael
Spira\cite{Prospino_stop}.}}
\end{center}
\end{figure*}

It is crucial to note that within the approximation of neglecting
the $r$ terms and the D-terms, this coupling does not survive in
the maximal mixing scenario, it is proportional to $\sin(
4\theta_{\stop})$. Nonetheless because of it its Yukawa nature
this can be a rather large coupling and therefore phase-space
allowing $Br(\stt \ra \sto h)$ can be large. Considering that
$\stt$ pair production exceeds $1$pb for $\mstt\leq 500$GeV (See
Fig.~\ref{t2t2_lhc}), $\stt$ can trigger Higgs (h)
production\footnote{$\sto \stt$ is completely negligible at the
LHC\cite{Prospino_stop}.}.

Contrary to $\chi_2^0 \ra \chi_1^0 h$ whose branching ratio can
reach $100\%$ for some of the SUSY parameters\cite{Baer_htochi}
and thus very efficiently triggers Higgs production, $Br(\stt \ra
\sto h)$\cite{ATLAS_SUSYtoh} can never reach $100\%$.
This is because, independently of other decay modes into the $b,
\tilde{b}$ sector, there is always the competing larger decay rate
$\stt \ra \sto Z$. Indeed when the splitting is large $\stt \ra
\sto Z$ can be approximated by $\stt \ra \sto \phi^0$, $\phi^0$
being the neutral Goldstone Boson, with an effective coupling $g
\frac{1}{4 \mw} \sin 2\theta_{\stop} \; (\msttsq-\mstosq)=g/2M_W
m_t (A_t+\mu/\tgb)$.

When $M_A$ is small, $\stt$ can also provide a welcome source of
pseudo-scalar (and heavy Higgses) through $\stt \ra \sto A,H$.
What's more, the strength of the $\stt \ra \sto A$ coupling does
not depend on the stop mixing angle:

\beqn \label{stop1stop2A}
V_{\sto \stt A}&=& i g  \frac{m_t}{2\mw}
\left(\frac{A_t}{\tgb}-\mu \right)
\eeqn

The decay $\stt \ra \sto H$ is generally smaller and vanishes when
the mixing is maximal. In the decoupling limit this becomes:
\beqn
\label{stop1stop2H} V_{\sto \stt H}& \sim & i g \cos 2\theta_{\stop} \frac{m_t}{2\mw}
\left(\frac{A_t}{\tgb}-\mu \right)
\eeqn

Of course to calculate the branching ratios of $\stt$ into Higgses
we have evaluated all possible widths of $\stt$, without QCD
corrections though. We have checked our numbers against those of
\cite{Bartl_stopdecays} as well as the output of {\tt GRACE}
\cite{GraceSusy}. For a general recent review of stop decays see
\cite{Stops_Porod,stop_Kraml}. For further reference note that
whenever stop mixing is not excessively small, we can reach
$Br(\stt \ra \sto h) \sim 10\%$. Associated $\stt \sto A$ in
mSUGRA has also been entertained recently\cite{Moretti_stophiggs}.
However in the mSUGRA scenario the mixing is generally not large
and the stops are usually heavy leading to small cross section for
Higgs production through stops. But then in this same scenario
large drops in the inclusive production due to stop mixing hardly
occur either.

\subsection{Constraints from low Higgs masses, $\Delta\rho$ and CCB}

Large values of the $\sto \sto h$ vertex which lead to the largest
$pp \ra \sto \sto h$ and the sharpest drop in $R_{gg\gamma\gamma}$
occur when the mixing is large with a large splitting between the
two stop physical masses. It is, however, for this configuration
that one has some strong constraints which preclude the highest
values of the cross section. For instance, one has to be wary that
imposing a lower bound on the Higgs mass, from its non observation
at LEP2 say, can restrict drastically the $\sin 2
\theta_{\stop}-\mstt$ parameter space. This constraint is very
much dependent on $\tgb$. Much less dependent on $\tgb$ but a
quite powerful one, for the values of $\msto$ that we have
entertained, is the constraint coming from $\Delta
\rho$\cite{Drhosusy}. Taking the present limit $\Delta \rho<.0013$
applicable to New Physics with a light
Higgs\cite{Langacker_Fits98}, which here means essentially the
contribution from stops and sbottoms (and marginally the Higgs
sector\footnote{For light stops in the decoupling limit the
sbottom-stop contribution when substantial gives a positive
contribution, whereas the Higgs sector contributes a negligible
negative contribution.} ) generally excludes region of the
parameter space where the $\sto \sto h$ is largest. In our $\Delta
\rho$ constraint we have relied on the two-loop calculation of
\cite{Drhosusy_2loop}, which can enhance $\Delta \rho$ by as much
as $10\%$ even with a heavy gluino.

One more constraint one needs to mention. In the stop sector and
in the presence of large mixing as is the case here, one often has
to check whether the parameters do not induce colour and charge
breaking global minima (CCB)\cite{CCBnaive}.
%
It has been argued that the constraints based on the global minima
may be too restrictive\cite{CCB_Kusenko}. It was shown that for a
wide range of parameters, the global CCB minimum becomes
irrelevant on the ground that the time required to reach the
lowest energy state exceeds the present age of the universe.
Taking the tunneling rate into account results in a milder
constraint which may be approximated\cite{CCB_Kusenko} by :

\beqn
\label{CCBkusenko}
A_t^2 +3\mu^2 < 7.5 (M_{\tilde{Q}_3}^2 + M_{\stop_R}^2)
\eeqn

When presenting our results we will, unless otherwise stated,
impose the limits $m_h>90$GeV, $\Delta \rho <.0013$ together with
the mild CCB constraint Eq.~\ref{CCBkusenko}. Considering that the
CCB constraint is rather uncertain, it is worth pointing out that
our CCB constraint hardly precludes points which are not already
rejected by $\Delta \rho$ and $m_h$.

Apart from the indirect constraints we also imposed, the model
independent limit, $\msto,\msbo>80$GeV from present direct
searches\cite{stoplimit98}. Our limit on the stop, is however
superseded by our constraint that the lightest neutralino is the
LSP and that $\sto \ra c \chi_1^0$\cite{stoptochi0} is always
open. When taking $\mu=-M_2=250$GeV with the unification
condition, $\chi_1^0 \simeq 120$GeV and thus $\msto
>120$GeV.

\subsection{$\tgb=2.5$}
We start our analysis by considering the case with $\tgb=2.5$.
Although this value is not far from being excluded by the direct
LEP2 searches\cite{Latest_mh_limit}, depending on the exact SUSY
parameters, we study it here in order to compare our results with
those in\cite{AbdelStop_Hgg_Loops} and to show a feature which is
not present for higher values of $\tgb$.
\subsubsection{The case of a common mass in the third generation
squark sector}

We first revisit the case\cite{Kileng_mixing,AbdelStop_Hgg_Loops}
of allowing, at the electroweak scale, a common mass for all the
supersymmetric masses of the third generation squarks:
$\tilde{m}_{\tilde{Q}_3}=\tilde{m}_{\tilde{U}_{3R}}=
\tilde{m}_{\tilde{D}_{3R}}=\tilde{m}_{\tilde{3}}$. Taking a common
value for the SU(2) and U(1) masses shows that unless the
effective tri-linear term is negligible, $\tilde{A_t}\sim 0$, this
leads to $|\sin 2 \theta_{\stop}|=1$, see Eq.~\ref{s2t}. We note
that contrary to what is claimed in\cite{AbdelStop_Hgg_Loops} this
situation, although common for the first two generation of
squarks, occurs only in exceptional situations in a model such as
mSUGRA. Moreover in mSUGRA $A_t$ is controlled almost entirely by
$m_{1/2}$, the common gaugino mass, and thus would not be
excessively large\cite{RGE}. Leaving this aside, this assumption
helps keep the number of parameters to a minimum while
concentrating on the impact of mixing. To that effect we have set,
apart from the common third family scalar quark
$\tilde{m}_{\tilde{3}}$ which was allowed to vary in the range
$100-1000$GeV and $M_A=1$TeV, all other sfermion masses to
$500$GeV. Moreover we have assumed the unification condition for
the gaugino masses and set the Higgsino mass $M_2=-\mu=250$GeV. We
then scanned over $A_t$, $-1000 \leq A_t/(GeV)\leq 1000$ and
$\tilde{m}_{\tilde{3}}$. Note that since we are scanning over both
positive and negative values of $A_t$, some important mixing
effects sensitive to the sign of $A_t \times \mu$ are covered even
though we have fixed the sign of $\mu$. Among the $2. 10^4$
generated point for each $\tgb$ half passed all the constraints.

\begin{figure*}[htbp]
\begin{center}
\mbox{\epsfxsize=14cm\epsfysize=8cm\epsffile{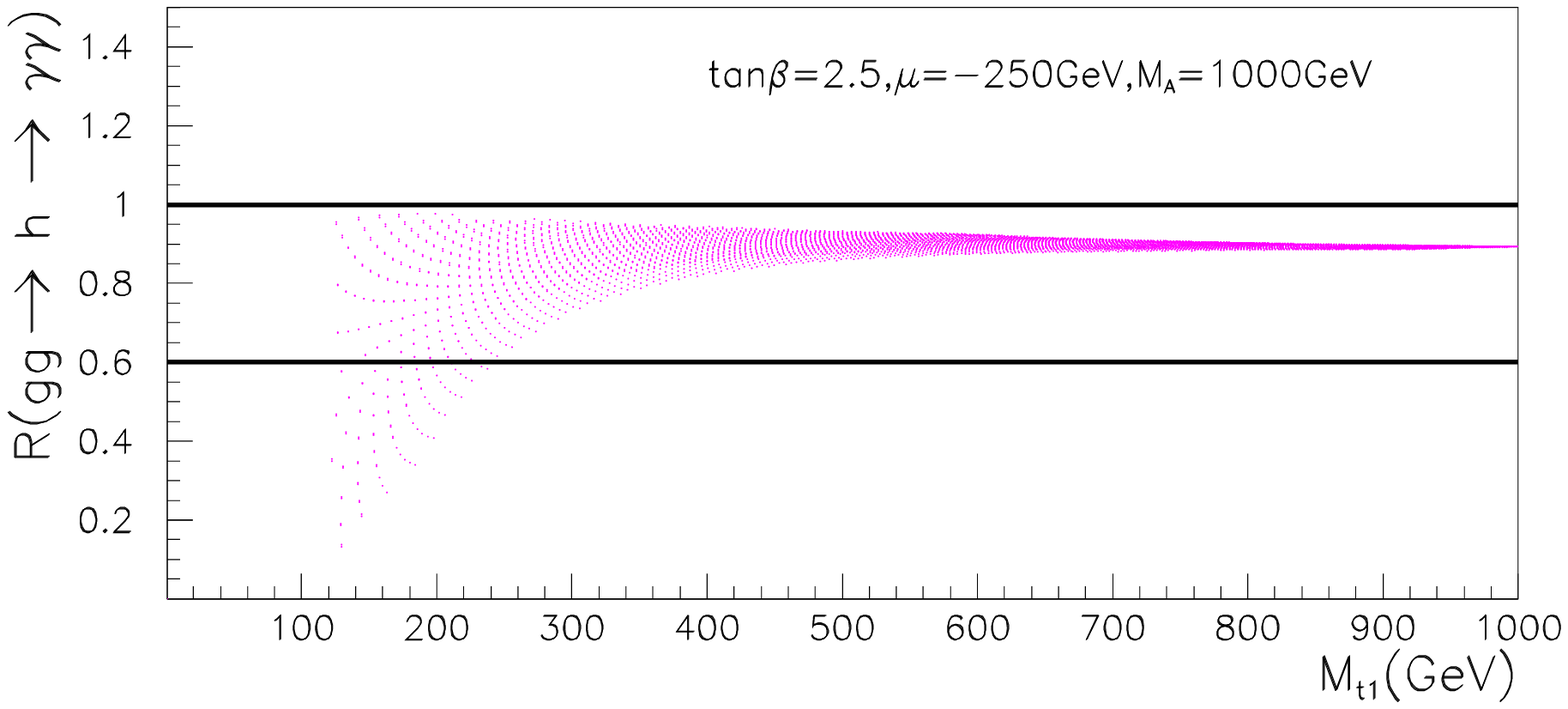}}
\caption{\label{tan25mu2501}{\em  $R_{gg\gamma\gamma}$ {\it vs}
$\msto$ for $\tgb=2.5, \mu=-250$GeV and $M_A=1$TeV. }}
\end{center}
\end{figure*}

First as shown in Fig.~\ref{tan25mu2501}, we do confirm that the
reduction in the two-photon signal in the direct channel is most
dramatic for the lowest values of the stop mass, although a low
stop mass does not always mean that a reduction has to occur. As a
matter of fact there are more points that generates a low $\msto$
and give $R_{gg\gamma\gamma}\ge .6$, say, than those that give
$R_{gg\gamma\gamma}\le .6$. Note that most points clustering
around values corresponding to little mixing or large stop masses.
Therefore the very rare situations corresponding to very sharp
drops could be interpreted as at best unnatural.
It is also worth pointing out that values such that
$R_{gg\gamma\gamma}\ge 1.$ are not obtained for $\tgb=2.5$. We
have verified that while, in principle, this was possible for
$\tgb=2.5$ this possibility was ruled out by the requirement of
having $m_h\ge 90$GeV.
\\

\begin{figure*}[p]
\begin{center}
\mbox{\epsfxsize=16cm\epsfysize=20cm\epsffile{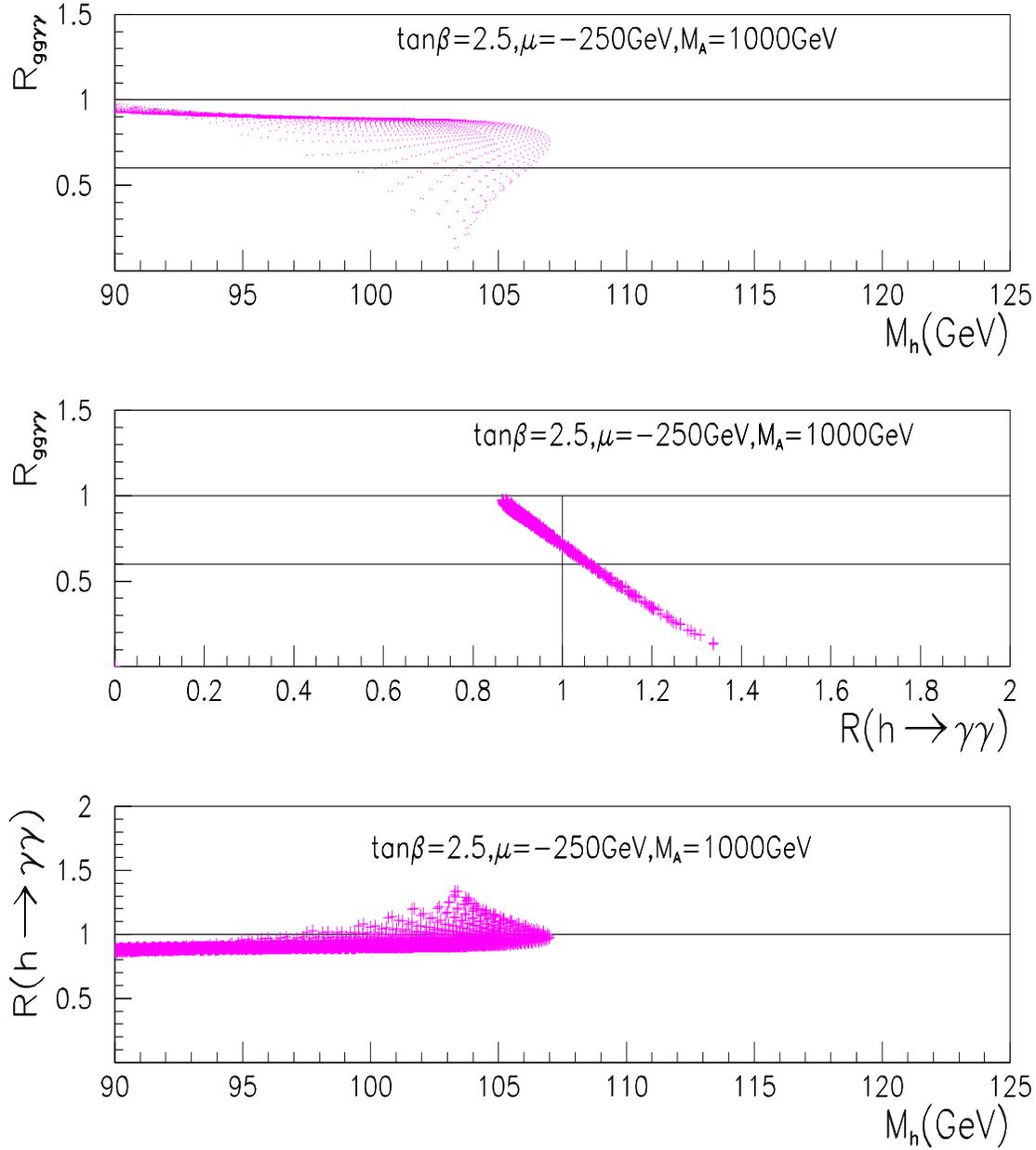}}
\caption{\label{tan25mu2502}{\em As in Fig.~\ref{tan25mu2501} but
for a) $R_{gg\gamma\gamma}$ {\it vs} $M_h$, b)
$R_{gg\gamma\gamma}$ {\it vs} $R_{\gamma\gamma}$ and c)
$R_{\gamma\gamma}$ {\it vs} $M_h$. }}
\end{center}
\end{figure*}

\noi As stressed numerous times, for the intermediate mass Higgs in the direct channel
decaying into two photons, the significance depends crucially on
the Higgs mass. It is therefore important to localise for which
values of the Higgs mass, the reductions are most drastic. For
$\tgb=2.5$ we see, Fig.~\ref{tan25mu2502}, that this reduction
gets worse, $R_{gg\gamma\gamma}\simeq .2$ for Higgs masses
clustered around $\sim 103$GeV.

It is important to note, on the other hand, that for Higgs masses
around $90$GeV where the (\sm) Higgs signal is most difficult to
extract, the effect of the stop is rather negligible (here there
is no mixing hence the low mass of the Higgs which does not get
further radiative corrections). Therefore this is a welcome point.
As compared to the case of CLASS-H with $M_A=180$GeV and
$\tgb=2.5$, for which $M_h=90$GeV, $R_{gg\gamma\gamma}$ reaches .3
whereas for the same Higgs mass (and $\tgb$) our points cluster
around one. An even more important remark concerns the behaviour
of the branching ratio into two photons. We find, see
Fig.~\ref{tan25mu2502}, that the branching ratio into photons in
this SUSY scenario increases at the same time as the direct
production decreases, in sharp contrast to what happens in CLASS-H
when $M_A$ decreases. This confirms our expectations.
The $R_{gg\gamma\gamma}$ {\it vs} $R_{\gamma \gamma}$ can be
considered as a signature of this scenario. In the corresponding
scatter plot of Fig.~\ref{tan25mu2502}, the points fall almost
along a line and shows that when $R_{gg\gamma\gamma}\le .8$,
$R_{\gamma \gamma}\ge 1$. Considering that in this large $M_A$
scenario and even in the presence of large mixing the $t \bar t h$
and $Wh$ are sensibly the same as in the standard model, the
associated Higgs production with the Higgs decaying into
two-photon should pose no problem with the high luminosity LHC. We
do not show the ratios for $t \bar t \gamma\gamma$ and $W\gamma
\gamma$ as these are given essentially by the ratio $R_{\gamma
\gamma}$, see Fig.~\ref{tan25mu2502} . To conclude, for this value
of $\tgb=2.5$, when $90\le m_h\le 100$ observability of the
lightest SUSY Higgs (h) is quite similar to that of the \smn.
Above these values, if the direct production is not possible, the
branching into photons is larger than the \sm and thus associated
production provides more chance of detecting the Higgs. For
instance, taking the \sm Higgs CMS analysis\cite{CMS_allHiggs}
with a luminosity of $100$fb$^{-1}$ as a guide, shows that it is
only in the range $100-105$ where values below .6 are possible for
$R_{gg\gamma \gamma}$ that the Higgs may not be observed in the
direct channel. The same analysis shows, however, that with the
values that we obtain in the associated channels that there is no
problem of cornering the Higgs. Note that for the most critical
drop in the direct channel we have obtained a enhancement factor
of up to $1.35$ in the associated production. For such values even
the ATLAS simulation\cite{Wh_tth_ATLAS,ATLAS_TDR} with a
luminosity of $100$fb$^{-1}$ indicates observation in the
associated channels.
\begin{figure*}[htb]
\begin{center}
\mbox{\epsfxsize=14cm\epsfysize=8cm\epsffile{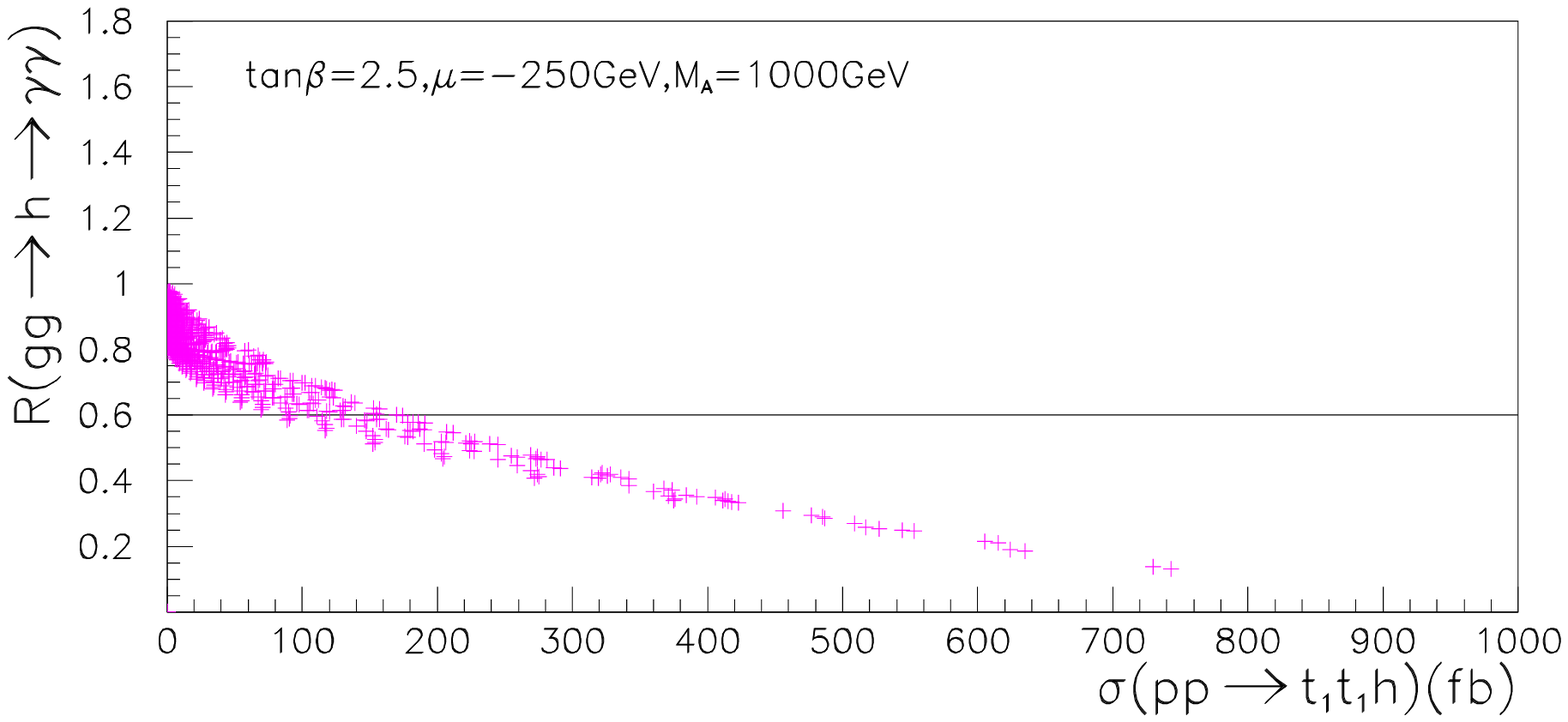}}
\caption{\label{tan25mu2503}{\em  As in Fig.~\ref{tan25mu2501} but
for $R_{gg\gamma\gamma}$ {\it vs} $\sigma(\sto \sto h)$(fb) .
 }}
\end{center}
\end{figure*}

Finally, another note of optimism in the case where the drop in
the inclusive production is severe is that production of $h$ in
association with stops could help also. As shown in
Fig.~\ref{tan25mu2503}, whenever $R_{gg\gamma\gamma} \le .6$,
$\sigma(\sto \sto h)$ is in excess of $100$fb and can reach as
much as $\sim 740$fb. As a comparison, for these extreme cases for
which $m_h\sim 100-105$GeV, one has $\sigma(t \bar t h)\simeq
500$fb. Considering that, see Fig.~\ref{tan25mu2501}, these
helpful $\sto \sto^* h$ cross sections are for values of $\msto
\le 250$GeV for which $R_{gg\gamma\gamma} \le .6$, $\sto$ with our
choice of parameters will decay exclusively into $c \chi_1^0$. It
remains to be seen whether this constitutes a viable signal and
whether we could use the Higgs decays into $b\bar b$, which by the
way is not much affected at these low values $\tgb$ by these
mixing effects. The signal would be $ b\bar b + {\rm jets} +
\slashp_T$. Note that the continuum $\sto \sto Z$ is quite small.
For $\msto=120$GeV and maximal stop mixing angle, after folding
with $Br(Z \ra b\bar b)$ the continuum leads to a dismal raw cross
section of about $1$fb.

\subsubsection{Lifting the degeneracy in the third family scalar masses}
\begin{figure*}[htbp]
\begin{center}
\mbox{\epsfxsize=15cm\epsfysize=15cm\epsffile{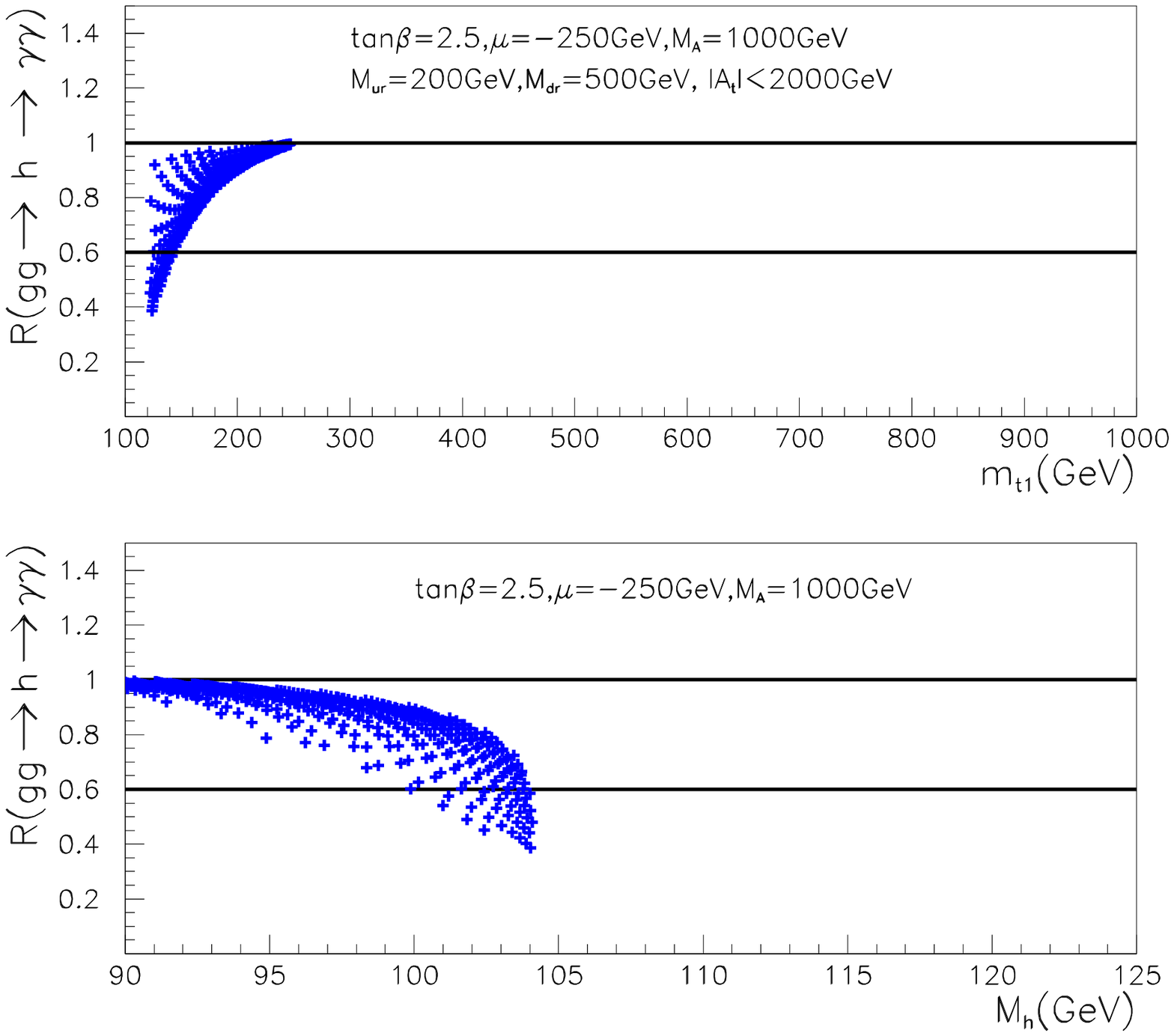}}
\caption{\label{noteq1at25}{\em  a) $R_{gg\gamma\gamma}$ {\it vs}
$\msto$ for $\tgb=2.5, \mu=-250$GeV and $M_A=1$TeV, when we allow
different scalar masses for the third generation as given, see
text . b) As in a) but for $R_{gg\gamma\gamma}$ {\it vs} $M_h$. }}
\end{center}
\end{figure*}

\begin{figure*}[htbp]
\begin{center}
\mbox{\epsfxsize=15cm\epsfysize=15cm\epsffile{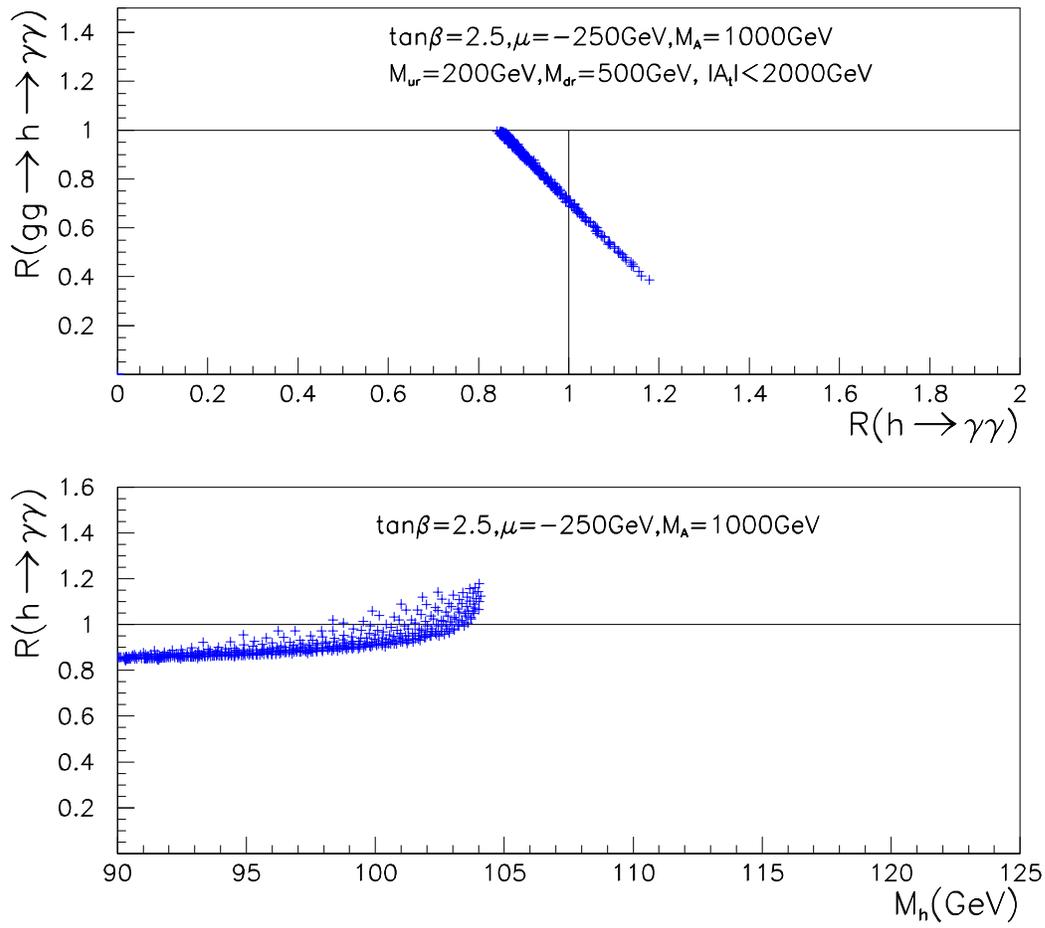}}
\caption{\label{noteq2at25}{\em  As in Fig.~\ref{noteq1at25} but
for $R_{gg\gamma\gamma}$ {\it vs} $R_{\gamma\gamma}$ and
$R_{\gamma\gamma}$ {\it vs} $M_h$ .  }}
\end{center}
\end{figure*}

\begin{figure*}[p]
\begin{center}
\mbox{\epsfxsize=16cm\epsfysize=20cm\epsffile{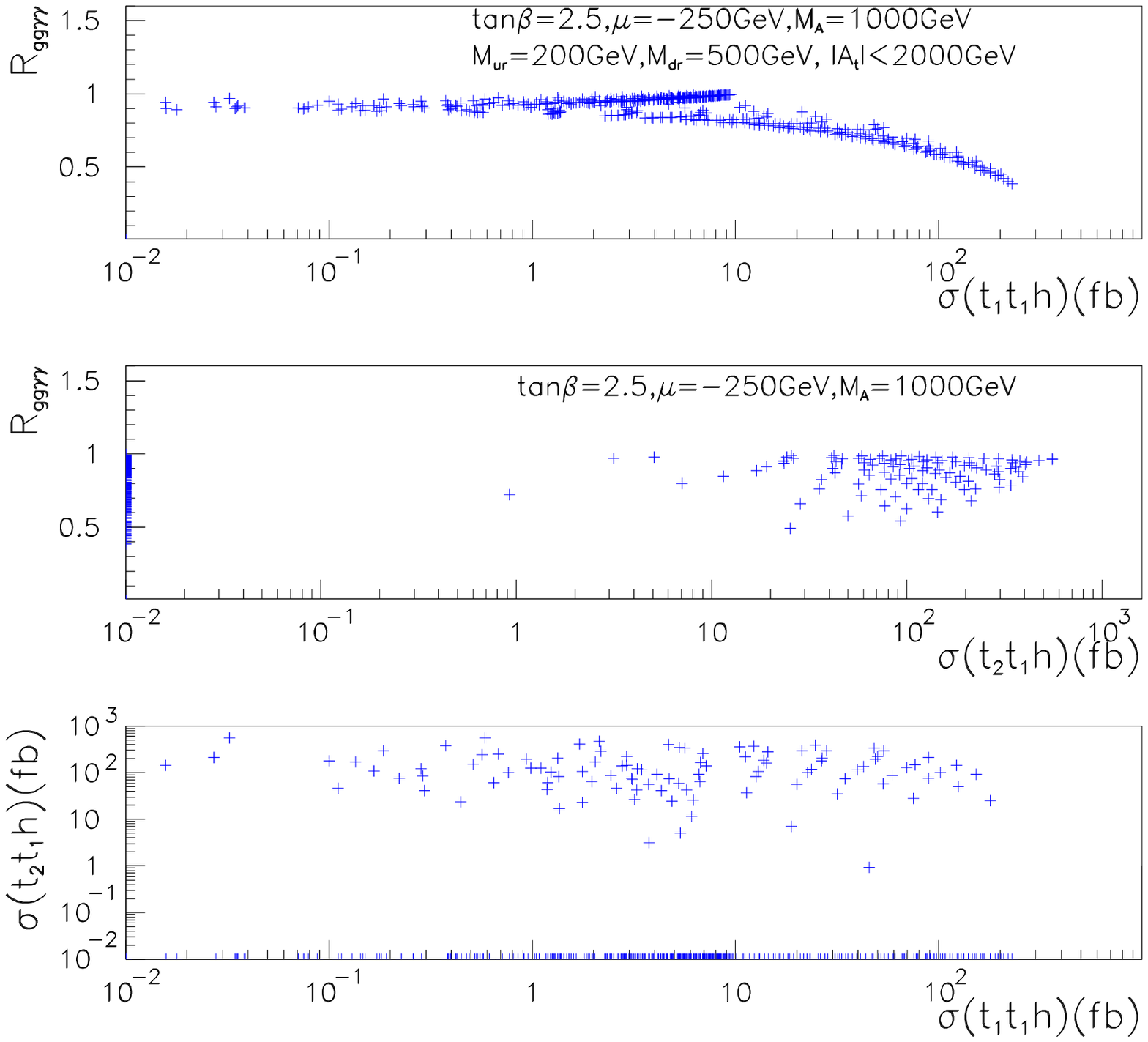}}
\caption{\label{noteq3at25}{\em  As in Fig.~\ref{noteq1at25} but
for a) $R_{gg\gamma\gamma}$ {\it vs} $\sigma(pp \ra \sto \sto h)$,
b)  $R_{gg\gamma\gamma}$ {\it vs} $\sigma(pp \ra \stt \sto h)$ and
 c) $\sigma(pp \ra \sto \sto h)$ {\it vs} $\sigma(pp \ra \stt
\sto h)$. $\sigma(pp \ra \stt \sto h) \equiv \sigma(pp \ra \stt
\sto^* h + \stt^* \sto h )$ }} 
\end{center}
\end{figure*}

We have already argued that the scenario with exactly equal squark
masses for the third generation is very special and even
unnatural. Taking a more general framework, we move away from the
case of maximal mixing. As we have discussed this can open up new
possibilities, notably $\stt \ra \sto h$ decays. For illustration,
we have taken $\tilde{m}_{\tilde{t}_{3R}}=200$GeV,
$\tilde{m}_{\tilde{b}_{3R}}=500$GeV and allowed $50 \le
\tilde{m}_{\tilde{Q}_3}\le 500$GeV. In order to compensate for the
deviation from maximal mixing, the trilinear coupling was allowed
to vary in the range $-2000\le A_t \le 2000$GeV. However very few
points with $|A_t|\ge 1200$ pass our constraints, essentially from
$\Delta \rho$. As expected the general features found in the case
of maximal mixing are still present here, even though with our
parameters the drops are not as dramatic as in the maximal mixing
case. Another observation is that $m_h>105$GeV is not generated.
This is because contrary to the previous case the stop masses do
not extend to $1$TeV and hence the radiative corrections to the
Higgs mass are not optimal. Nonetheless as seen in
Fig.~\ref{noteq1at25} a ratio $R_{gg\gamma \gamma}$ as low as .4
is possible and occurs for low $\sto$ masses. Again this drop
occurs for a small range of Higgs masses sensibly the same as in
the maximal mixing case, $m_h\sim 103-104$GeV,
Fig.~\ref{noteq1at25}. However when this occurs one is saved by
the fact that the branching ratio into photons is larger than in
the \smn, Fig.~\ref{noteq2at25}. Moreover we still find that when
$R_{gg\gamma\gamma}$ gets too small $pp \ra \sto \sto^*h$ is of
the order $100$fb reaching a maximum of $200$fb when
$R_{gg\gamma\gamma}$ is lowest, Fig.~\ref{noteq3at25}. The main
novelty here is $pp \ra \stt \stt^* \ra \stt \sto^* h$, with
$Br(\stt \ra \sto h)={{\cal O}}(10\%)$. Because this stems from a
two-body cross section, it can lead to quite large $\sigma(\stt
\sto^* h)$ reaching as much $600$fb, and therefore in many
instances larger than the continuum $\sto \sto^* h$,
Fig.~\ref{noteq3at25}. What is also worth noting is that these
large cross sections do not necessarily occur when one has large
drops in the inclusive two-photon channel. Moreover the signature
in this channel should be cleaner, taking advantage of the cascade
decays of the other $\stt$ starting with $\sto Z, \sbo W, b
\tilde{\chi}^+_{1,2},...$. Of course there are points where
neither $\stt \sto h$\footnote{With our set of parameters one
would expect that some points with maximal mixing are generated.
However we have checked that these do not pass all the
constraints. This explains why we never get a vanishingly small
$\stt \sto h$ cross section.} nor $\sto \sto h$ exceeds
$10$fb,Fig.~\ref{noteq3at25}. However in this case the reduction
in $R_{gg\gamma \gamma}$ is quite modest.

\subsubsection{Stop mixing with a low $M_A$}

\begin{figure*}[htbp]
\begin{center}
\mbox{\epsfxsize=16cm\epsfysize=20cm\epsffile{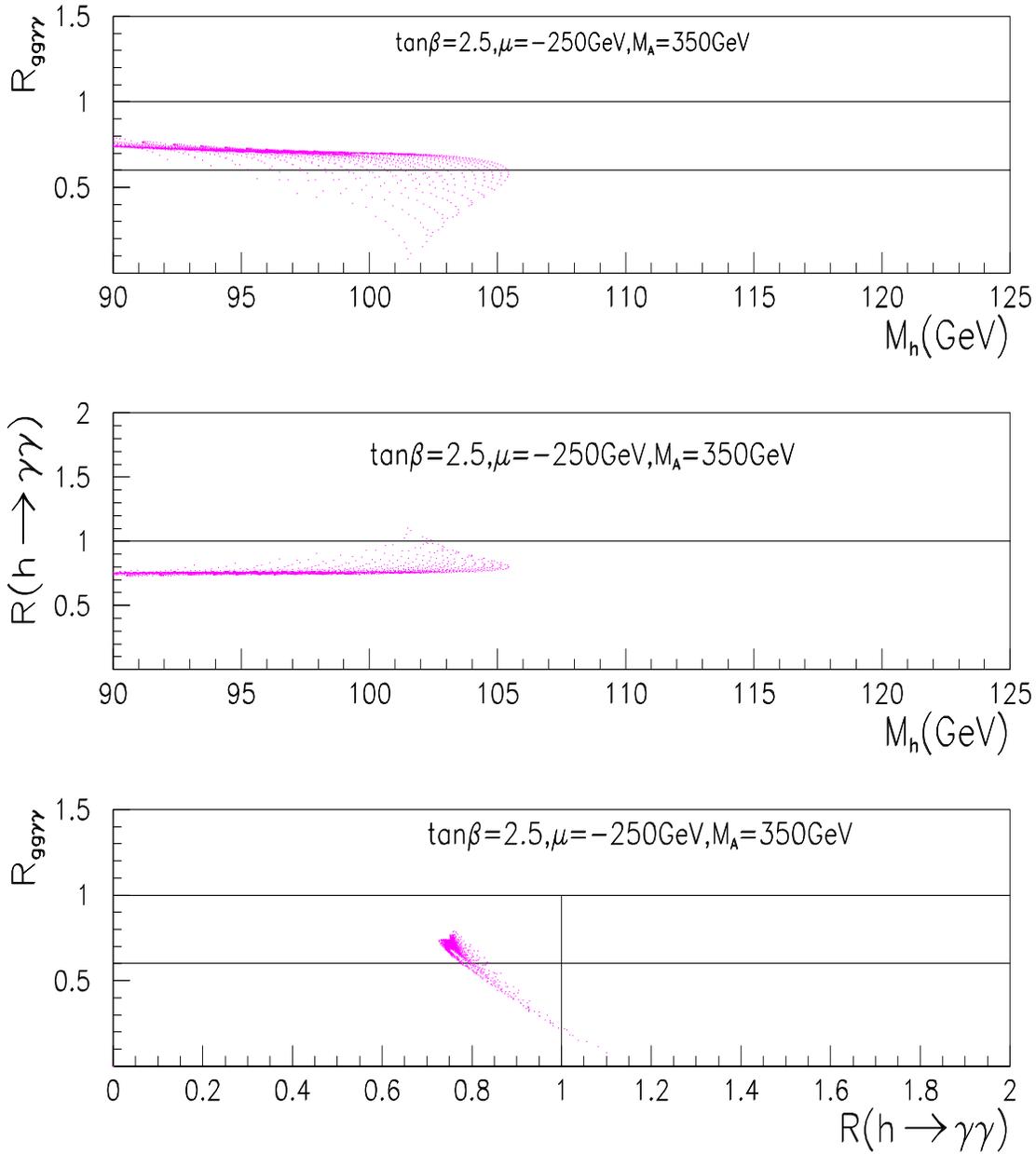}}
\caption{\label{tan25a3501}{\em \em a) $R_{gg\gamma\gamma}$ {\it
vs} $M_h$ for $\tgb=2.5, \mu=-250$GeV and $M_A=350$GeV, with equal
squark masses. b) As in a) but for $R_{\gamma\gamma}$ {\it vs}
$M_h$.
 c) As in a) but for $R_{gg\gamma\gamma}$ {\it vs}
$R_{\gamma \gamma}$. }} 
\end{center}
\end{figure*}

\begin{figure*}[htbp]
\begin{center}
\mbox{\epsfxsize=14cm\epsfysize=8cm\epsffile{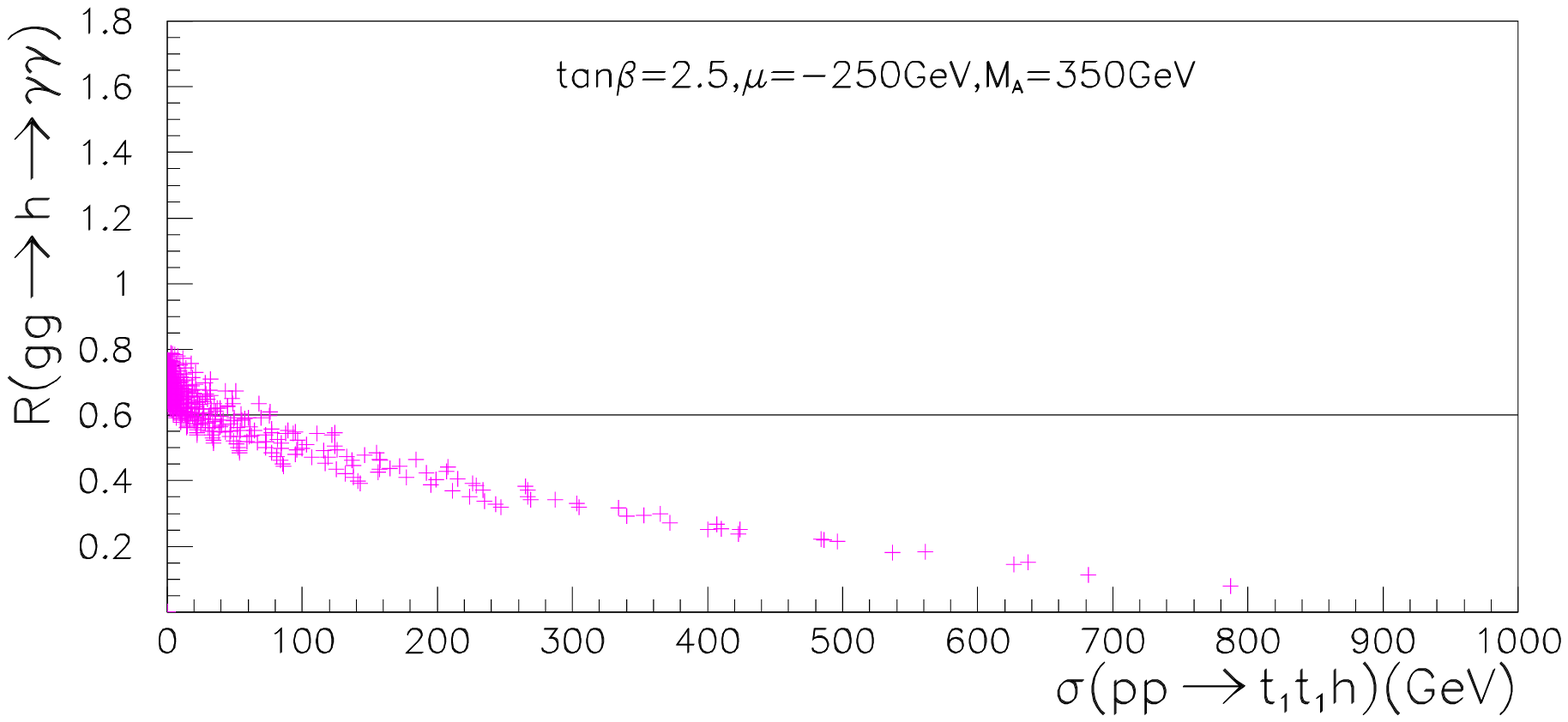}}
\caption{\label{tan25a3502}{\em As in
Fig.~\ref{tan25a3501} but for $R_{gg\gamma\gamma}$ {\it vs}
$\sigma(pp \ra \sto \sto h)$ .}}
\end{center}
\end{figure*}

We have seen in section 2, in the case of no-mixing, that as $M_A$
decreases both the inclusive and associated two-photon channels
decrease, mainly because of an increase in the width into $b\bar
b$ which dominates the total width and hence reduces the
two-photon branching ratio. Since $h b \bar b$ is hardly affected
by the mixing effect, this decrease due to $M_A$ will also be
present in the case of mixing and hence reduces the significance
of the two-photon channel. This overall reduction, independent of
mixing, can be evaluated by using Eq.~\ref{RggmA}. Likewise since
the $\sto \sto h$ vertex carries the same reduction $R$ as the $t
\bar t h$ vertex, see Eq.~\ref{stopstophcoupling2}, an overall
$M_A$- reduction in $R_{gg\gamma \gamma}$, which can be
approximated by Eq.~\ref{RglgluggmA}, will take effect beside the
pure large stop mixing effects that we have discussed in the
$M_A=1$TeV limit. We first consider the case of a moderate
$M_A=350$GeV with all other masses as in section 3.5.1.
Figs.~\ref{tan25a3501} show that those points for which at large
$M_A$ the effect of mixing were most drastic on
$R_{gg\gamma\gamma}$ are not much further reduced. They occur for
masses which are sensibly the same as with the much larger $M_A$.
Note however that the optimal values of $R_{gg\gamma\gamma}$ are
reduced from about 1 to .8 and occur also for $M_h \sim 90$GeV.
This reduction is essentially what we would have obtained by
applying the factor $R_{gg\gamma\gamma}$ calculated using
Eq.~\ref{RglgluggmA}. For this value of $M_h$ now detectability
may be a problem if the luminosity is low, especially that the
corresponding $R_{\gamma \gamma}$ is about .75, which may also
preclude detection in the associated Higgs production,
Figs.~\ref{tan25a3501}. With these values occuring at such low
values of $m_h$, even CMS\cite{CMS_allHiggs} with $30fb^{-1}$ will
miss the Higgs, but again there should be no problem in the
associated production after collecting $\sim 100fb^{-1}$. Still,
whenever mixing becomes important and reduces $R_{gg\gamma
\gamma}$ significantly, associated production should be no
problem. For instance when $R_{gg\gamma \gamma}$ is below .4,
$R_{\gamma \gamma}\ge .95$.
In these configurations $Br(h \ra \gamma\gamma)$ benefits from the
increase in $\Gamma (h \ra \gamma \gamma)$ which is not completely
offset by the increase in $\Gamma(h \ra b \bar b)$. In these
configurations with small $\sto$, $\sto \sto h$ could help with
$\sigma(\sto \sto h)=100-780$fb, Fig.~\ref{tan25a3502}.

\begin{figure*}[p]
\begin{center}
\mbox{\epsfxsize=14cm\epsfysize=14cm\epsffile{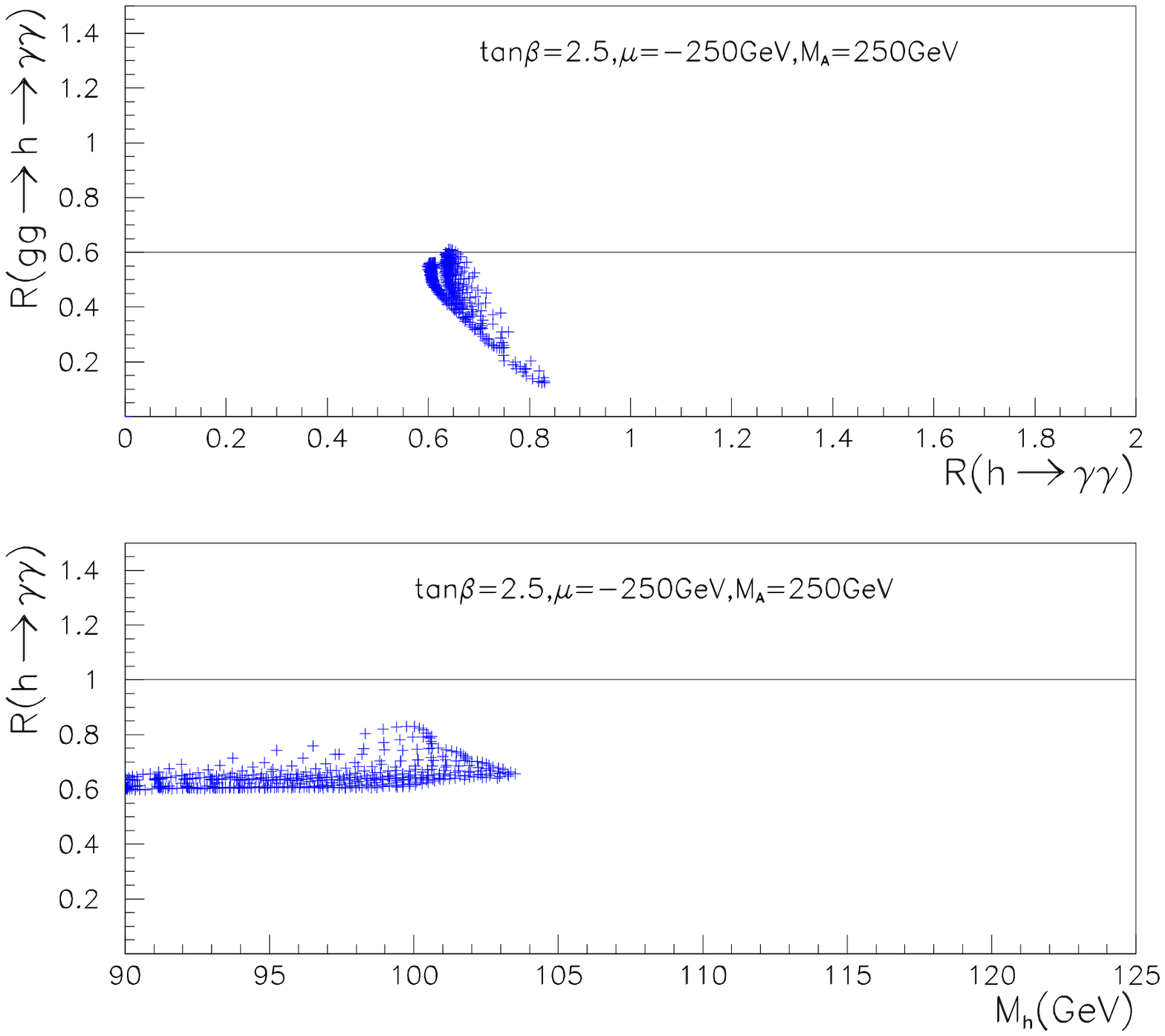}}
\caption{\label{tan25ma2501}{\em \em a) $R_{\gamma\gamma}$ {\it
vs} $M_h$ for $\tgb=2.5, \mu=-250$ and $M_A=250$TeV, with equal
squark masses. b) As in a) but for $R_{gg\gamma\gamma}$ {\it vs}
$R_{\gamma \gamma}$}}
\end{center}
\end{figure*}
\begin{figure*}[p]
\begin{center}
\mbox{\epsfxsize=14cm\epsfysize=14cm\epsffile{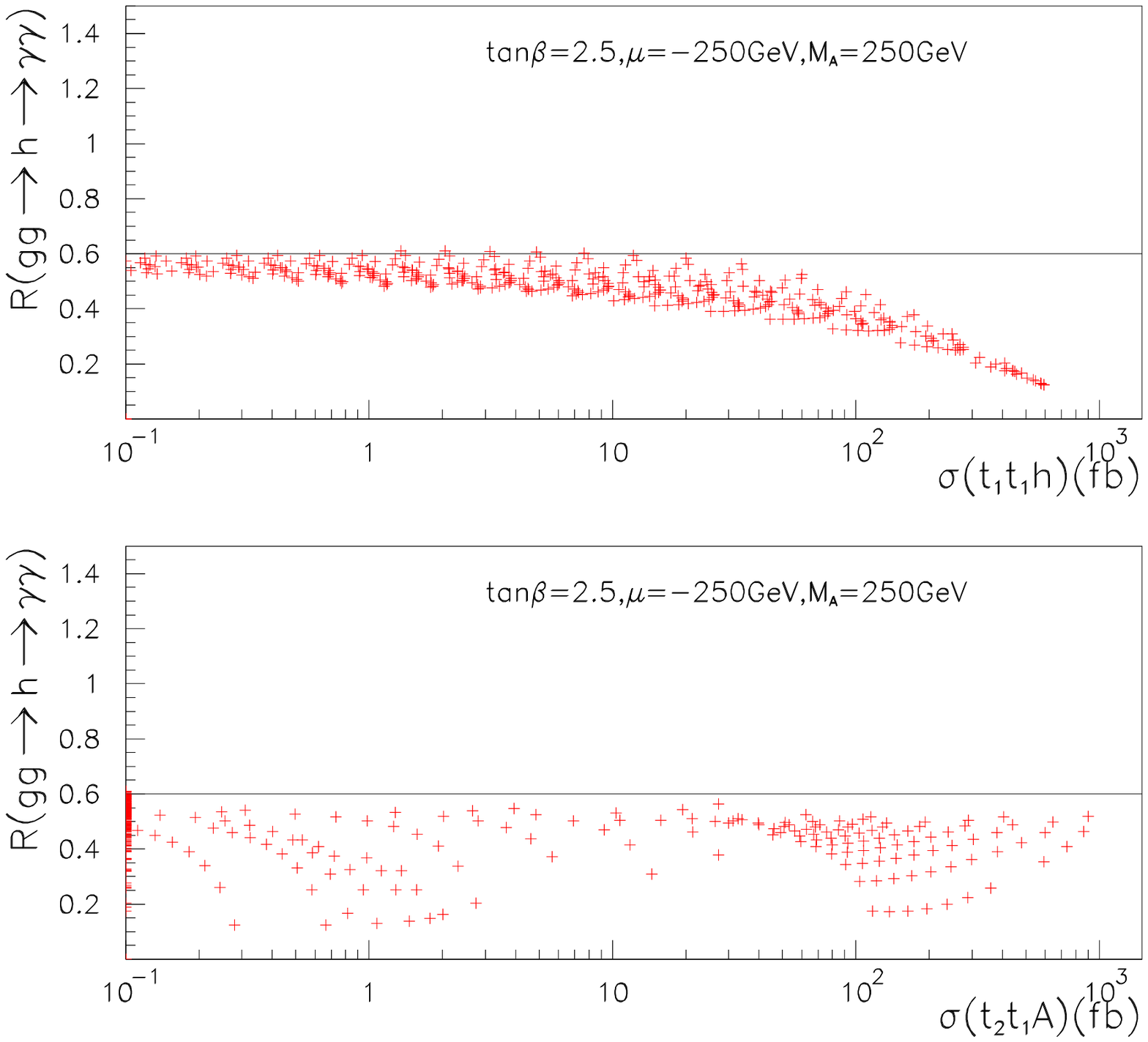}}
\caption{\label{tan25ma2502}{\em  As in Fig.~\ref{tan25ma2502} but
for $R_{\gamma\gamma}$ {\it vs} $\sigma(\sto \sto h)$ and
$\sigma(\stt \sto A)$.  }} 
\end{center}
\end{figure*}

We may argue that had we taken a much lower value of $M_A$ we
would have introduced a larger reduction in $R_{\gamma \gamma}$
which may affect dangerously the associated production. We would
then be in a situation where the inclusive cross section is down
because of large mixing in the stops and the associated production
small mainly because the branching into photons is down as a
result of $M_A$ being low. Note however that in these situations
we would be far from  the decoupling regime, with all Higgses
being relatively light and a very light stop having large
couplings to the Higgses. One consequence of this light spectrum
is that, even in the case of maximal stop mixing where $\stt \ra
\sto h$ is inhibited, the $\sto \stt A$ coupling is large,
Eq.~\ref{stop1stop2A} and can be such that it triggers $\stt \ra
\sto A$. This is because large mixing and large splitting between
the stop, allows enough phase space for a relatively light
pseudo-scalar. To illustrate this fact, we have lowered $M_A$ to
250GeV. The gross features found for $M_A=350$GeV are still
present as concerns the inclusive production of $h$
Fig.~\ref{tan25ma2501}, with an overall reduction factor due to
$Br(h \ra \gamma \gamma)$ which is slightly larger. At the same
time the location of the drops are shifted to slightly lower
values of $m_h$, which is a direct consequence of a low $M_A$.
However as shown in Fig.~ \ref{tan25ma2502}, $\sigma(\stt \sto A)$
can be quite large and often exceeds $\sto \sto h$. Note that
$\sigma(\stt \sto^* A)$ may be large even for points where the
inclusive two-photon cross section is lowest, whereas $\stt \sto^*
h$ is largest for regions where the inclusive cross section is
most affected. Therefore we see that combining different channels
in this scenario offers much better prospects than in the
no-mixing case with the same low value of $M_A$. To start with,
when the direct production is very much reduced, associated
production has a better siginificance in the case of very large
stop mixing compared to the no mixing case for the same $M_A$.
Another interesting point is that although the main decay of $A$
will be into $b \bar b$, we also find that $A \ra Zh$ can be
substantial. For instance, the chain $\sigma(pp \ra \stt \stt \ra
\stt \sto A \ra \stt \sto Zh)$ can reach as much as $350$fb (for
this point $\msto=129$GeV, $\mstt=396$GeV). For larger values of
the stop masses ($\msto=235$GeV, $\mstt=525$GeV), the same chain
corresponds to $43$fb. The decay $\stt \ra \sto H$ is also
possible, but the corresponding cross section, $\sigma(\stt
\sto^*H)$  is below $10$fb. This is because the branching ratio
into $H$ is about a factor $\cos^2 2\theta_{\stop}$ down compared
to the branching ratio into $A$, while $H$ and $A$ are almost
degenerate in mass, Eqs.~\ref{stop1stop2H}-~\ref{stop1stop2A}. To
end this section let us mention that when the mass of the
pseudo-scalar gets small, below $2 m_t$, one should also
investigate direct $g g \ra A,H$ production. A low mass $\sto$ has
no effect either the production or decay (we are in scenario where
$\mstt >M_A$) of $A$, the usual channels should not be much
affected.  For $H$, one needs to critically review how the
production is affected and whether $H \ra \sto \sto$ can be
exploited. The phenomenology is certainly richer here and the
Higgs(es) should not be missed.

\subsection{$\tgb=5$}
We now move to a larger $\tgb$. We go through basically the same
steps as those in the previous section, 3.5. For the same
scenarios we will scan over the same mass ranges. One general new
feature will have to do with the fact that for larger $\tgb$ we
obviously have larger Higgs masses. In most cases this will help.
However on the whole similar conclusions will be reached.

\subsubsection{The case of a common mass in the third generation
squark sector with large $M_A$}

\begin{figure*}[p]
\begin{center}
\mbox{\epsfxsize=14cm\epsfysize=14cm\epsffile{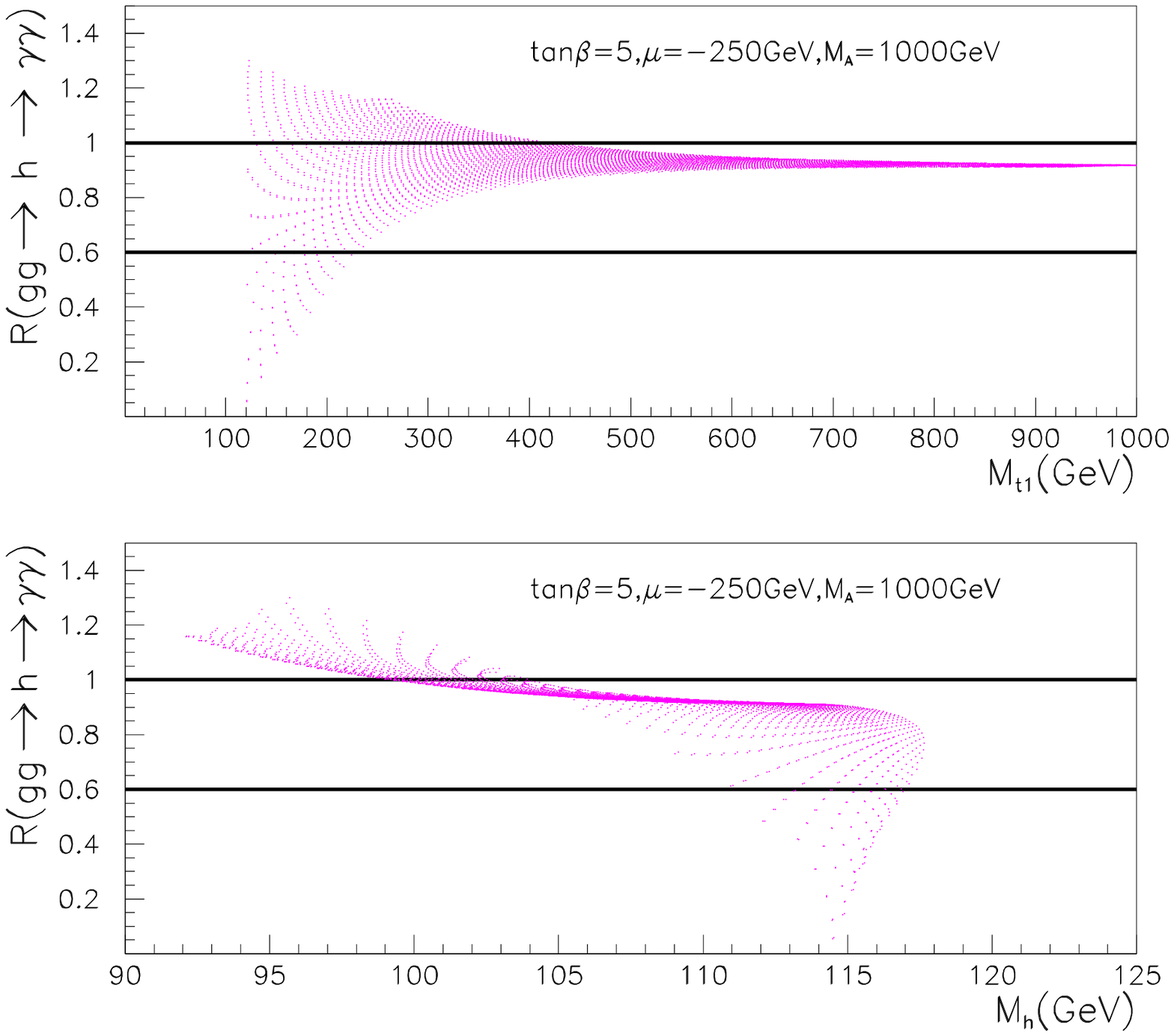}}
\caption{\label{tan5x1}{\em  a) $R_{gg\gamma\gamma}$ {\it vs}
$\msto$ for $\tgb=5, \mu=-250$GeV and $M_A=1$TeV. b) As in a) but
for $R_{gg\gamma\gamma}$ {\it vs} $M_h$ .}}
\end{center}
\end{figure*}

\begin{figure*}[p]
\begin{center}
\mbox{\epsfxsize=14cm\epsfysize=14cm\epsffile{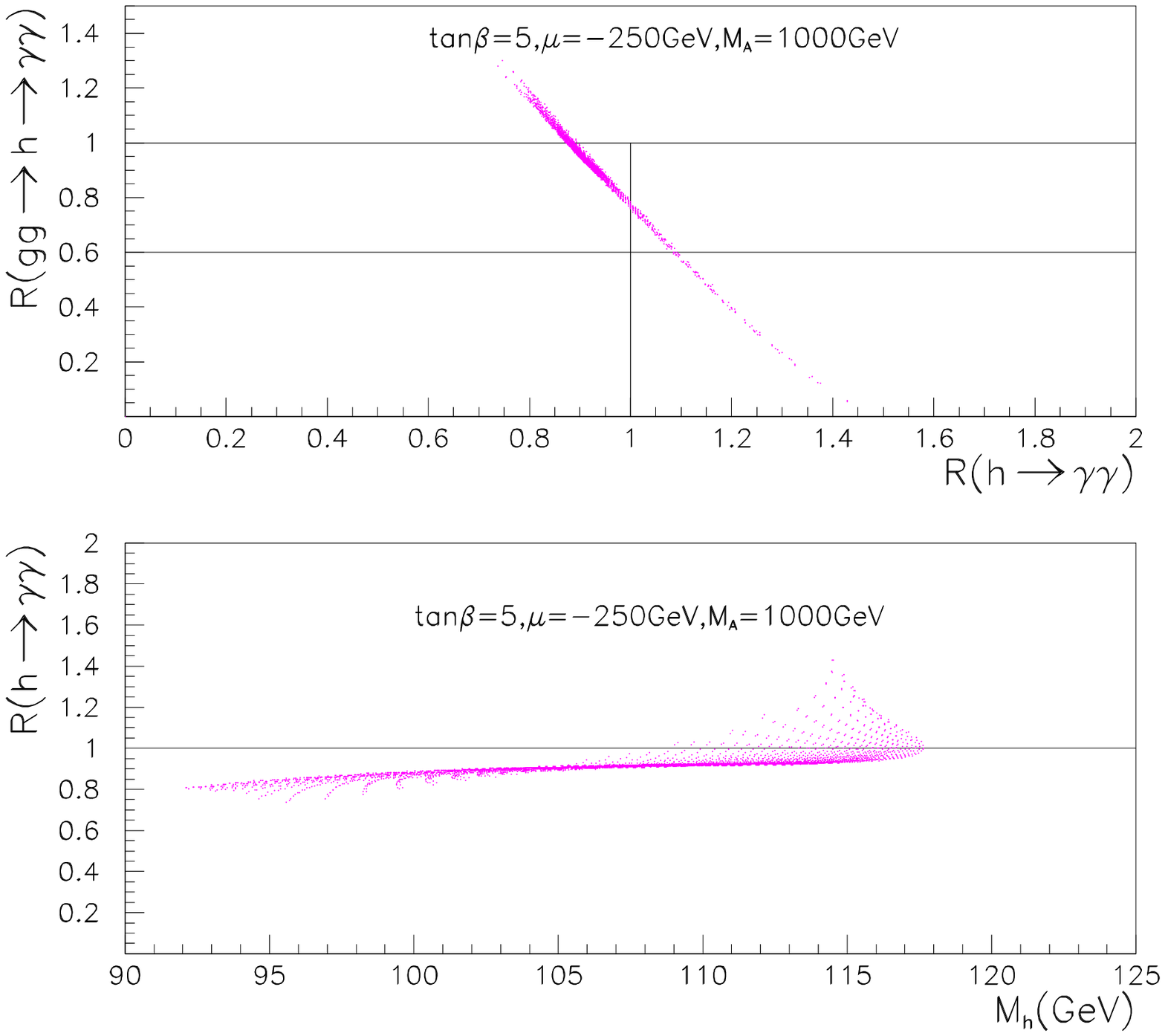}}
\caption{\label{tan5x2}{\em a) As in Fig.~\ref{tan5x1} but for
$R_{gg\gamma\gamma}$ {\it vs} $R_{\gamma\gamma}$.    b) As in a)
but for $R_{\gamma\gamma}$ {\it vs} $M_h$ .}}
\end{center}
\end{figure*}

\begin{figure*}[htbp]
\begin{center}
\mbox{\epsfxsize=14cm\epsfysize=8cm\epsffile{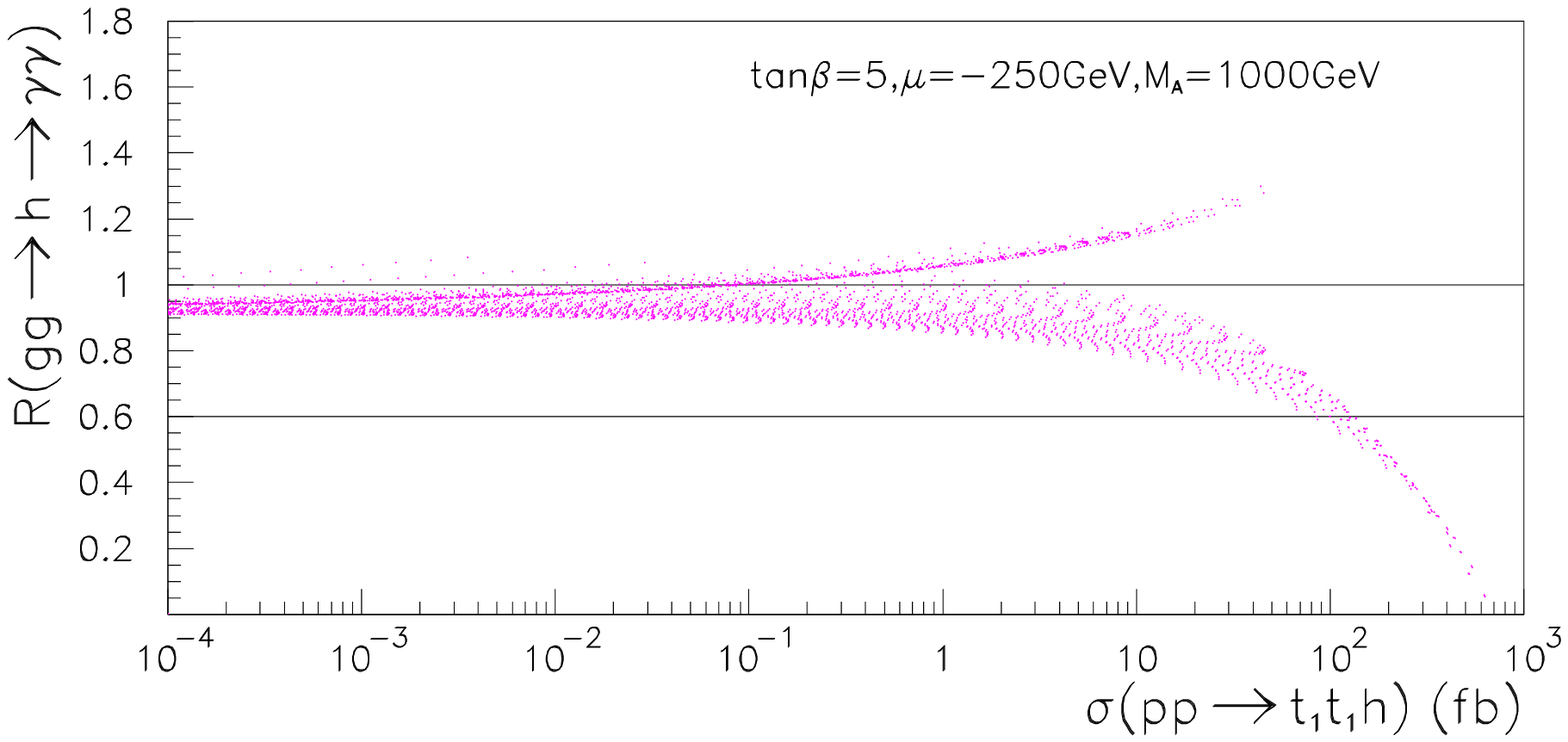}}
\caption{\label{tan5x3}{\em  As in Fig.~\ref{tan5x1} but for
$R_{\gamma\gamma}$ {\it vs} $\sigma(\sto \sto h)$.  }}
\end{center}
\end{figure*}

Again $R_{gg\gamma \gamma}$ is most affected when the $\sto$ mass
is smallest, Fig.~\ref{tan5x1}. In the maximal mixing case, one
new feature compared to $\tgb=2.5$ is that the ratio $R_{gg\gamma
\gamma}$ can be larger than one, for small $\msto$, reaching
almost $\sim 1.3$. This is even more welcome that it occurs for
Higgs masses in the range $92-98$GeV, Fig.~\ref{tan5x1}. As a
matter of fact, this is consistent with the argument we gave
earlier: in this case $\stt$ is not too heavy so that the top and
$\sto$ loop interfere and since the scale in the stop sector is
not too high, the radiative corrections to the lightest Higgs mass
are far from maximal. Considering that, especially in the lower
end of this range, the significance in the direct channel are
usually (\sm or no-mixing) smallest, such scenarios can make it
easier to discover $h$ even in the direct channel. Of course,
light $\sto$ (with much heavier $\stt$) can also lead to a much
reduced $R_{gg\gamma \gamma}$. When this happens it occurs for
higher Higgs masses, clustered around $M_h=115$GeV. Though for
this range of $m_h$ significances in the direct production are
much better, for certain values of the parameters the drop is too
severe: $R_{gg \gamma \gamma} < .4$. But again this occurs
simultaneously with an enhanced $R_{\gamma \gamma}$: $R_{\gamma
\gamma}>1.2$, Fig.~\ref{tan5x2}. Again the smaller $R_{gg \gamma
\gamma}$ the larger $R_{\gamma \gamma}$. As with the lower $\tgb$
when the direct production drops, $\sigma(pp \ra \sto \sto h)$
increases. When $R_{gg\gamma\gamma}<.6$ this cross section is in
excess of $100$fb up to $\simeq 650$fb, for the smallest value of
$R_{gg\gamma\gamma}$, Fig.~\ref{tan5x3}. Note also that when
$R_{gg \gamma \gamma}>1$ this additional cross section is below
$100$fb.

\subsubsection{The case of a common mass in the third generation
squark sector with  $M_A=350$GeV}

\begin{figure*}[p]
\begin{center}
\mbox{\epsfxsize=14cm\epsfysize=14cm\epsffile{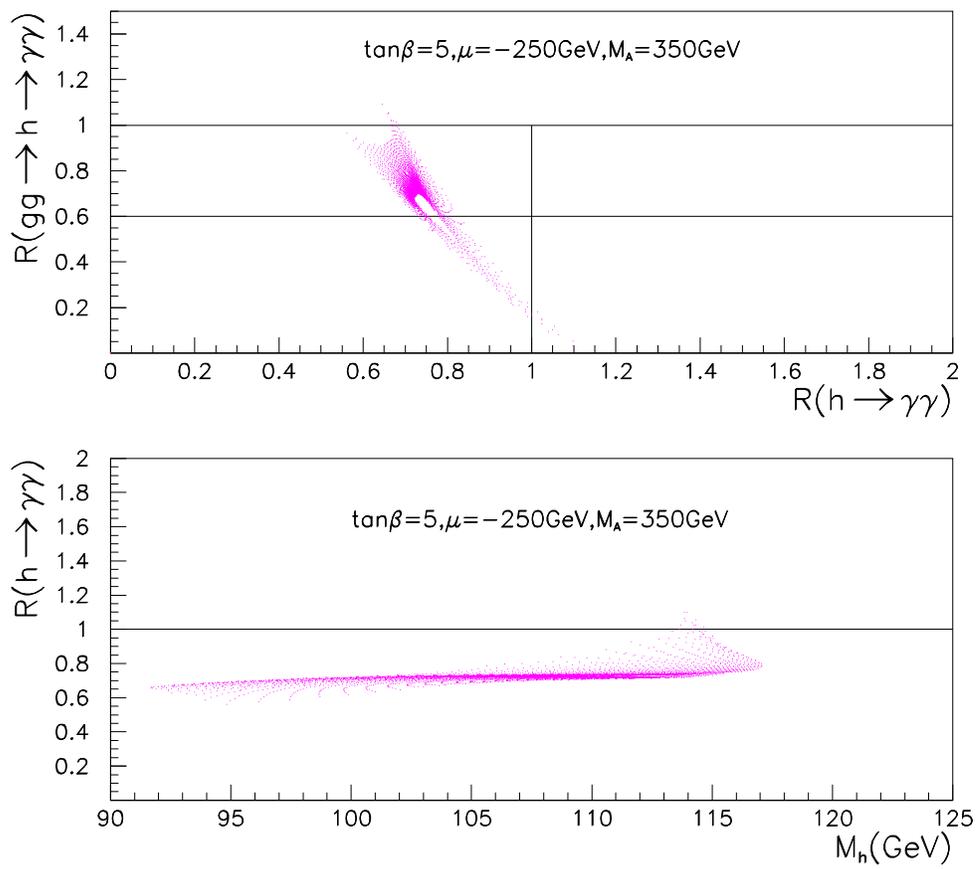}}
\caption{\label{tan5x5}{\em a) As in Fig.~\ref{tan5x2} but for
$R_{gg\gamma\gamma}$ {\it vs} $R_{\gamma\gamma}$.    b) As in a)
but for $R_{\gamma\gamma}$ {\it vs} $M_h$ .}}
\end{center}
\end{figure*}

\begin{figure*}[htbp]
\begin{center}
\mbox{\epsfxsize=14cm\epsfysize=7cm\epsffile{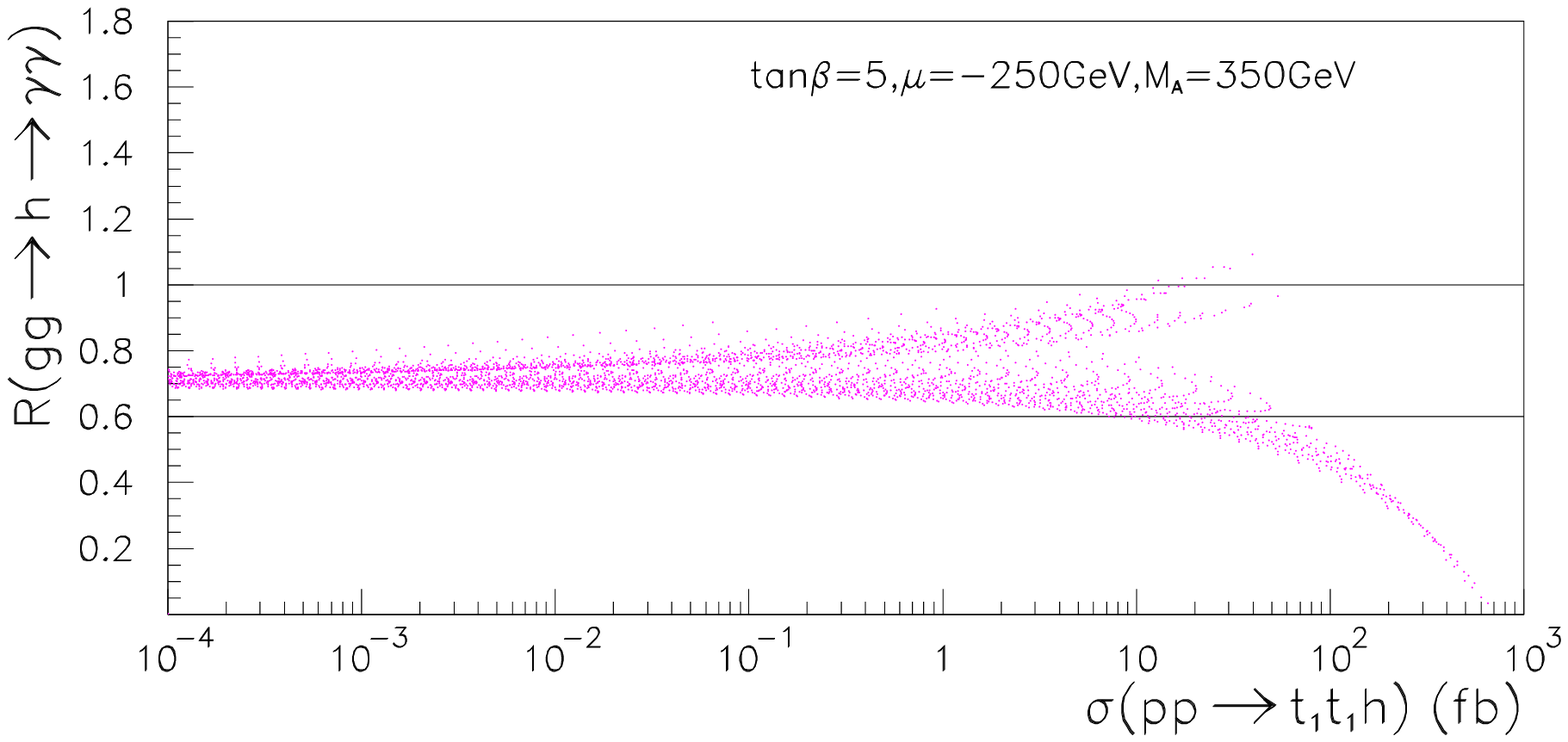}}
\caption{\label{tan5x6}{\em  As in Fig.~\ref{tan5x1} but for
$R_{\gamma\gamma}$ {\it vs} $\sigma(\sto \sto h)$.  }}
\end{center}
\end{figure*}

The discussion is essentially the same as the one we presented for
$\tgb=2.5$ with $M_A=350$GeV. The overall reduction factor from
the lowering of $M_A$ which affects $R_{\gamma \gamma}$ is slighty
smaller (about .76) but then the reductions in $R_{gg \gamma
\gamma}$ are for $m_h\sim 115$GeV, Fig.~\ref{tan5x5}.   Note also
that  $\sigma(\sto \sto h)$ production, Fig.~\ref{tan5x6}, is only
slightly smaller than with  $\tgb=2.5$ (this is due to a higher
Higgs mass) and therefore is a useful addition when the direct
channel drops too much. For $R_{gg \gamma \gamma}<.2$ one gets as
much as $~400$fb. For larger $\tgb$ and small $M_A$, de-excitation
of $\stt$ into $\sto$ is not as efficient as for the lower $\tgb$
with the rather moderate values of $\mu$ that we have considered
in this study. This is evident from Eq.~\ref{stop1stop2A}, but as
we see $\sto \sto h$ still plays its role.

\subsubsection{Lifting the degeneracy in the third family scalar masses}
\begin{figure*}[p]
\begin{center}
\mbox{\epsfxsize=14cm\epsfysize=14cm\epsffile{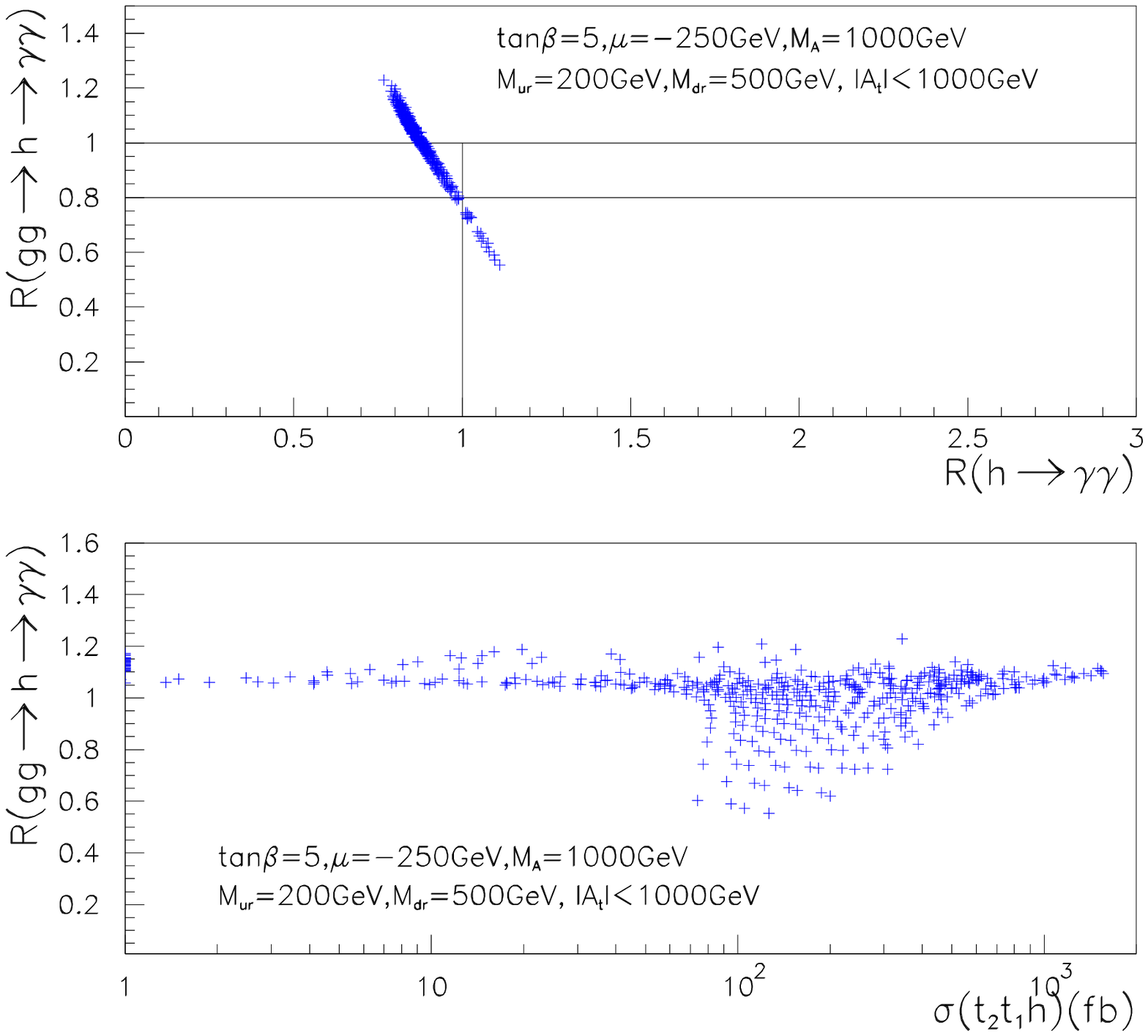}}
\caption{\label{tan5x7}{\em \em a) $R_{gg\gamma\gamma}$ {\it vs}
$R_{\gamma \gamma}$  for $\tgb=5, \mu=-250$GeV and $M_A=1$TeV,
when we allow different scalar masses for the third generation as
given, see text . b) As in a) but for $R_{gg\gamma\gamma}$ {\it
vs} $\stt \sto h$}}
\end{center}
\end{figure*}

\begin{figure*}[p]
\begin{center}
\mbox{\epsfxsize=14cm\epsfysize=14cm\epsffile{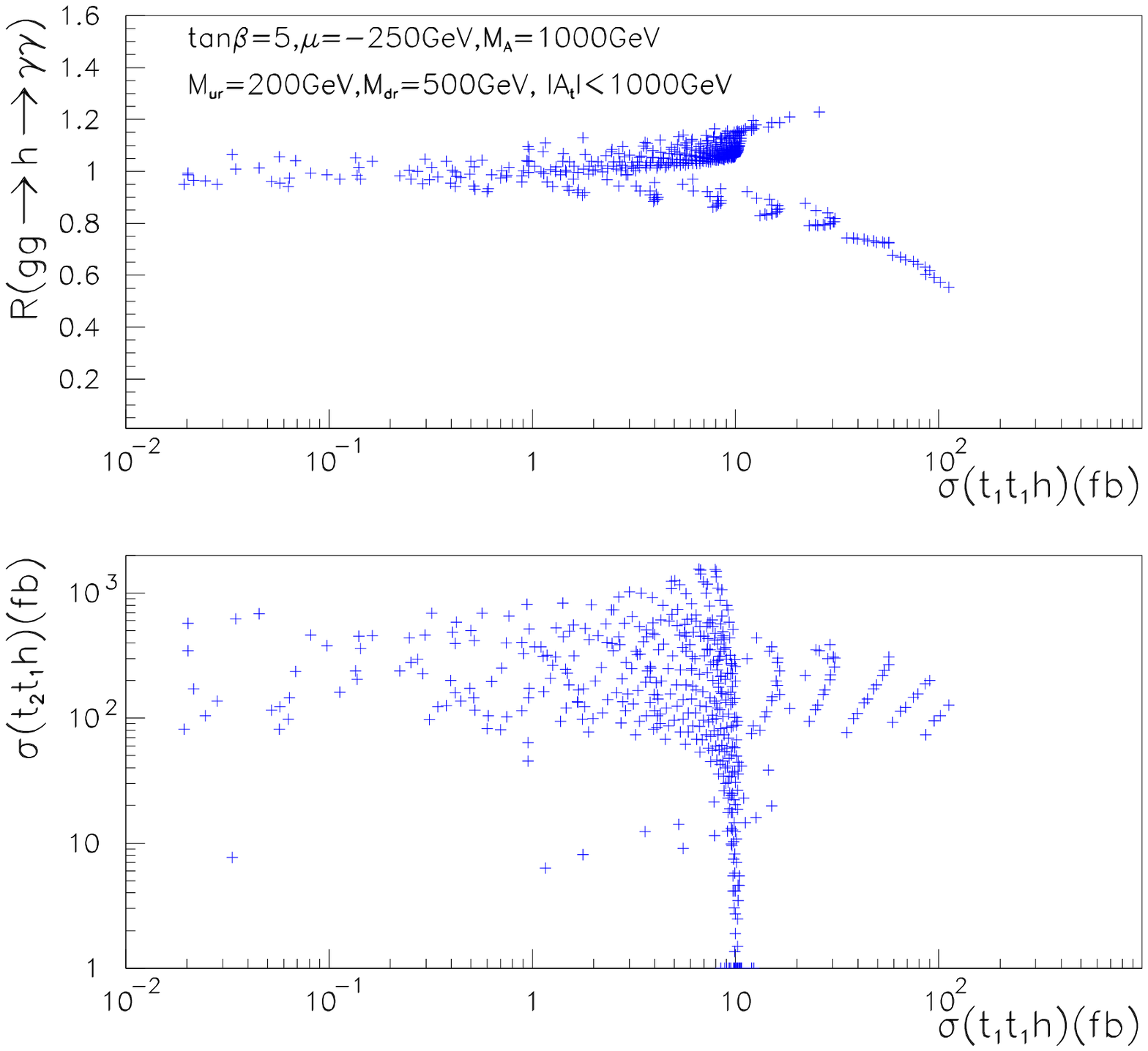}}
\caption{\label{tan5x8}{\em  As in Fig.~\ref{tan5x7} but for
$R_{gg\gamma\gamma}$ {\it vs} $\sto \sto h$ and $\stt \sto h$ {\it
vs} $\stt \sto h$ .  }}
\end{center}
\end{figure*}

Taking unequal masses as in section 3.5.3, the reductions in the
direct production are less pronounced. We do not get below $R_{gg
\gamma \gamma}<.55$, while values up to 1.25 are still possible
for $R_{gg \gamma \gamma}$. There is also little change in where
these reductions or enhancements occur as a function of the Higgs
mass. Again when $R_{gg\gamma \gamma}<.8$, $R_{\gamma \gamma}>1$,
Fig.~\ref{tan5x7}. As with $\tgb=2.5$ when the mass degeneracy is
lifted, the channel $\stt \ra \sto h$ opens up.This leads to
typical cross sections of the order of $100$fb especially for
regions where the drop in the direct inclusive two-photon channel
is the largest, see Fig.~\ref{tan5x7}. This cross section can be
larger  than $1$pb in situation where $R_{gg \gamma \gamma}$ is
little reduced. Of course our scans do show some regions where
this cross section is unusable, typically when the mixing and stop
splitting is small, but then as Fig.~\ref{tan5x7} shows the
inclusive two-photon channel is unaffected. Of course continuum
$\sigma(\sto \sto^* h)$ is still useful when large drops occurs
(it is then around $100$fb) but note that for our choice of
parameters $\sigma(\stt \sto^* h)$ is practically always larger,
Fig.~\ref{tan5x8}.

\subsection{$\tgb=10$}
\begin{figure*}[p]
\begin{center}
\mbox{\epsfxsize=16cm\epsfysize=20cm\epsffile{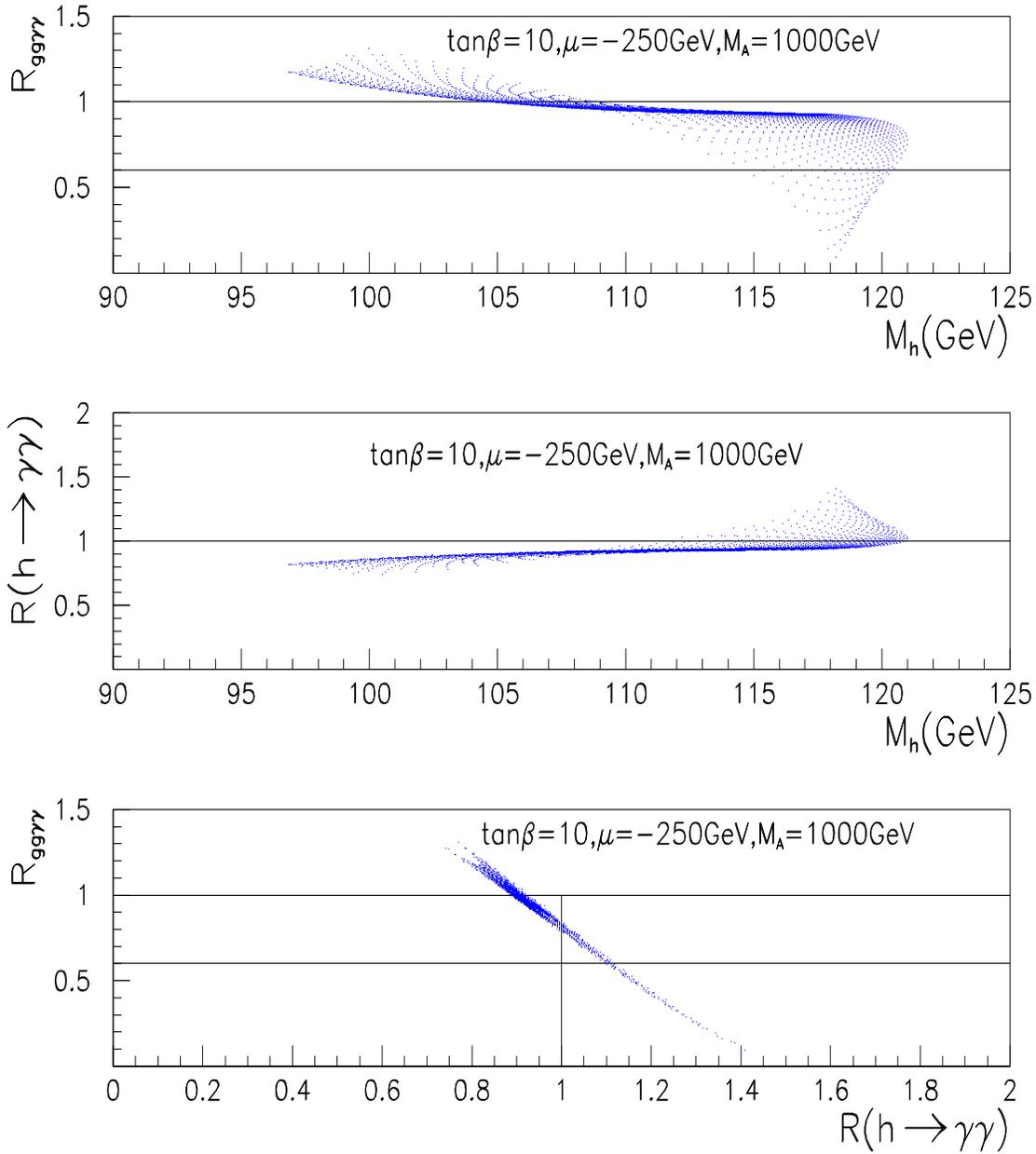}}
\caption{\label{tan10x1}{\em \em a) $R_{gg\gamma\gamma}$ {\it vs}
$m_h$ for $\tgb=10, \mu=-250$GeV and $M_A=1$TeV, b) As in a) but
for $R_{h\ra \gamma\gamma}$ {\it vs} $M_h$ , c) As in a) but for
$R_{gg\gamma\gamma}$ {\it vs} $R_{\gamma\gamma}$ }}
\end{center}
\end{figure*}

Apart from the location, in terms of the Higgs mass, of where the
largest drop in $R_{gg\gamma\gamma}$ occurs, that is around
$118$GeV, all the general features we found in the case with
$\tgb=5$ are recovered again, Figs.~\ref{tan10x1}. Note that we do
not get more noticeable reduction either in $R_{gg \gamma \gamma}$
or $R_{\gamma \gamma}$ due to the larger $\tgb$, and also that
$R_{gg \gamma \gamma}>1$ are possible. Similar observations to
those made for $\tgb=5$ can be made here even when we consider
different splitting and lowering of masses, especially  as
concerns the importance of $\stt \sto^* h$. Some of  these results
are summarised in Figs.~\ref{tan10x1} .

\newpage
\section{Conclusions}
We  have in this paper reinvestigated the fate of the photon
signal of the lightest SUSY Higgs at the LHC when large tri-linear
mixing terms in the stop sector are present. Previous
investigations\cite{Kileng_mixing,AbdelStop_Hgg_Loops} had drawn a
very pessimistic picture of these scenarios. Our analysis shows
that if we exploit all the consequences of these scenarios and not
pick out only the Higgs signal in the inclusive channel these
models have an excellent discovery potential. First, the large
reductions in the inclusive two-photon signal not only require
large mixing but also that one of the stops be relatively light.
Although this has not been stressed in the text, a first signal of
these scenarios will be $\sto \sto^*$ production with a cross
section of order $\sim 100$pb.  Even though it may be argued that
regions with the largest drops in the inclusive two-photon channel
correspond to a very light stop and are likely to lead to a
signature, jets+$\slashp_T$, which is difficult. In any case it
should be stressed that a hallmark of these scenarios is that
whenever the signal in the inclusive channel drops that in the
associated $Wh/Zh$ and $t\bar t h$ channels increases and makes it
up for the drop in the former channel. Moreover when $\sigma(pp
\ra h \ra \gamma \gamma)$ gets too small the continuum $pp \ra
\sto \sto h$ \cite{stophiggs_LHC} reaches values of order few
$100$fb. More importantly we find that since these situations
imply a large mass splitting between the two stops,  $\stt \ra
\sto h$ can be substantial leading to another source of Higgs
production with a yield larger than in the continuum and with a
better signature than the $\sto \sto^* h$ continuum. We have shown
that $\stt \ra \sto^* h$  occurs whenever the stop mixing angle
does not take its maximal value, $|\sin 2 \theta_{\stop}|=1$,
which is often unnaturaly assumed on the basis of equal soft SUSY
breaking masses for the $SU(2)$ and $U(1)$ sfermions of the third
generation, at the electroweak scale. We have also shown that
although when $M_A$ gets small the two-photon signals (both direct
and associated) get further reduced (this happens even in the
absence of mixing), with large tri-linear mixing terms and
especially for low values of $\tgb$, one can trigger $A$
production through the cascade $\stt \ra \sto A$, beside the usual
channels for $A$ productions. Moreover one should not forget that
especially with not too small $\tgb$, $\tgb >3$, scenarios with
light stops (but small mixing) do give an increase in the direct
channel, but then an decrease in the associated two-photon
channels. The overall conclusion we can draw almost resembles that
of a no-lose scenario: whenever an effect reduces a particular
signal it opens up new channels or enhances other channels. We
have not discussed the use of $h \ra b \bar b$ in the associated
$t \bar t h$ channel which in these scenarios should allow
detection. This requires rather good $b$-tagging facilities, as
shown in \cite{ATLAS_htobb}. This should certainly add to the
discovery potential. The new associated stop Higgs  signatures
deserve a full simulation to critically quantify how beneficial
these additions can be. In general there is a lack of detailed
study of stop phenomenology at the LHC despite some important
theoretical issues related to the third generation sfermions. As
has been pointed out by several authors \cite{Inverted_hierarchy}
the idea of an inverted hierarchy of the SUSY spectrum whereby the
third generation sfermions are, at the electroweak scale, much
lighter than the first two is compelling and quite plausible. This
helps solve the flavour problem in SUSY since very large masses
for the superpartners of the first two generations can suppress
FCNC, contributions to electric dipole moments and lepton flavour
violations. This would still not go against naturalness  since
these particles couple weakly to the Higgs, at the heart of the
fine-tuning problem. Naturalness does on the other hand require
the stops and sbottoms (and the electroweak gauginos higgsinos) to
be rather light, like in the scenarios we have studied and could
also with a light stop make electroweak
baryogenesis\cite{Stop_Baryogenesis} work.

\vspace*{1cm}

\noi {\bf Acknowledgments}

We would like to thank Guillaume Eynard for providing with his
programme for calculating the significance of the two-photon
associated Higgs signal using the ATLAS simulation and for very
useful discussions. We also thank Michael Spira for promptly
providing us with the code for the NLO stop pair production at the
hadron colliders and Andrei Semenov for advice on the use of {\tt
CompHep}. We are also grateful to Elzbieta Richter-Wa\c{s} for
helpful discussions and communication,as well as providing us with
an advance copy of the ATLAS TDR on Higgs Physics. K.S.
acknowledges the hospitality of LAPTH where part of this work was
done.  This work is done under partial financial support of the
Indo-French Collaboration IFCPAR-1701-1 {\em Collider Physics}.

\end{document}